# Author's note on arXiv submission

This is a slightly edited version of my PhD dissertation submitted and accepted in 2016. I have chosen to submit it to arXiv because it, as an overview article with a set of appended papers, will probably never be reworked into a book. Nevertheless I think it still has some merit – specifically the recent focus on design thinking and co-creation makes my focus of materials and participatory design in a pragmatist perspective timely. Also I regularly get requests on how and where to find the dissertation, and I figured it was worth having it online somewhere public-yet-not-secret (like my website would be).

The slight edits are mainly copyright-related: I am not allowed to republish ACM papers, so I have therefore attached the abstracts of the papers in their place, as well as included a link to the full paper. If you for some reason do not have ACM-access, send me a mail. The last paper, "(The Role of) Materials in Design Processes" is currently unpublished and on the backburner. I might try to publish it sometime, but for now, feel free to cite it as is, or just read it. It does constitute an interesting insight into the writing skills of an, at that point, almost finished PhD researcher, although I am more pleased with the write-up of the pragmatist perspective on materials that appears in the actual overview article.

The proper way to cite the actual dissertation would be (correct for your own reference format):

Nicolai Brodersen Hansen. 2016. "Materials in Participatory Design Processes". PhD Dissertation. Department of Culture and Communication, Aarhus University, Denmark.

<3

Nicolai Brodersen Hansen
nbhansen@gmail.com
Aarhus University, Denmark
March, 2017.

# MATERIALS IN PARTICIPATORY DESIGN PROCESSES

Nicolai Brodersen Hansen

PhD dissertation

# Materials in participatory design processes

A dissertation presented to the Faculty of Arts

Aarhus University

Denmark

by

Nicolai Brodersen Hansen

June, 2016

# Thanks to

Supervisor Kim Halskov, for careful guidance, vision, ambition and the human touch needed to make a phd a most pleasant journey.

co-supervisor Peter Dalsgaard for great discussions on pragmatism, for inspiring and challenging me in my time as a student and now as a colleague in CIBIS.

The research centres CAVI, Participatory IT (PIT), Creativity in Blended Interaction Spaces (CIBIS) and the Faculty of Arts for supporting and financing my project.

Our external partners at Dokk1, 3XN and BIG.

Ditte Basballe for being a good friend from we first joined forces as students in 2004 until now as colleagues and office mates.

Michael Mose Biskjær, Susanne Bødker, Clemens Klokmose, Siemen Baader and Nanna Inie in the CIBIS project, which transformed my phd work into something more coherent and focused.

Erik Stolterman, for inviting me to School of Informatics and Computing in Bloomington, Indiana, something I still rank as one of the best experiences of my entire phd.

The entire graduate cohort at the School of Informatics and Computing, for making me feel welcome and taking me to all those fantastic places. Especially thanks to Haley, Omar and Shad for being such fantastic human beings, as well as brilliant scholars.

The entire phd squad at our department, with a special shoutout to the VrUF-organisers: you made coming into the office something to be cherished.

All my students through the years, especially the three iterations of the course Advanced Interaction Design on Digital Design from years 2013-2015 - most days I learned more from you than you could ever imagine.

My colleagues and friends at Katrinebjerg, from the departments of Digital Design, Information Studies, Media Studies and Computer Science.

My family who I do not spend enough time with.

Inge, my love.

# Abstract

This dissertation presents three years of academic inquiry into the question of what role materials play in interaction design and participatory design processes. The dissertation aims at developing conceptual tools, based on Dewey's pragmatism, for understanding how materials aid design reflection.

It has been developed using a research-through-design approach in which the author has conducted practical design work in order to investigate and experiment with using materials to scaffold design inquiry. The results of the PhD work is submitted as seven separate papers, submitted to esteemed journals and conferences within the field of interaction design and HCI.

The work is motivated both by the growing interest in materials in interaction design and HCI and the interest in design processes and collaboration within those fields. At the core of the dissertation lies an interest in the many different materials used during the design process: sketches, prototypes as well as the materials we shape products out of: physical and digital materials now form a unity of computation and physical materials that has given rise to a new research interest in design and materiality.

The main results from the dissertation are an understanding of *design materials* that draws on pragmatist philosophy. The papers and overview article highlights how materials in a pragmatist perspective are more than the matter out of which we shape an idea. Rather they structure the entire process of inquiry, helping us frame problems, inspire solutions and try out these solutions in practice. This framework, developed in several of the submitted papers, is tested and illustrated through a series of experimental design cases.

# Opsummering

Denne afhandling præsenterer tre års akademisk undersøgelse af spørgsmålet om, hvilken rolle materialer spiller i interaktionsdesign og participatory design processer. Afhandlingen har til formål at udvikle konceptuelle værktøjer, baseret på Deweys pragmatisme, for at forstå, hvordan materialer består refleksion i designprocesser.

Undersøgelsen er blevet udført ved hjælp af en research-through-design tilgang, hvor forfatteren har gennemført praktisk design arbejde ved at undersøge og eksperimentere med at bruge materialer i eksperimentelle design cases. Resultaterne af ph.d.-arbejdet er udmøntet i syv artikler som er blevet afleveres som syv separate papirer, som er blevet optaget på velestimerede tidskrifter og konferencer inden for interaktionsdesign og HCI.

Arbejdet er motiveret både af den voksende interesse for materialer i interaktionsdesign og HCI, såvel som interessen for designprocesser og samarbejde inden for disse områder. Centralt i afhandlingen står en interesse for hvordan vi kan forstå brugen af de mange forskellige materialer, der anvendes i designprocessen: skitser, prototyper samt de materialer, vi former produkter ud af: fysiske og digitale materialer der i disse år smelter sammen til nye materialer der har computation som udgangspunkt. Derved dannes et nyt forskningsområde, og det er et designproces-blik på dette som afhandlingen tilbyder.

Afhandlingens hovedbidrag er en forståelse af *design materials,* baseret på pragmatisk filosofi. Artiklerne og kappen fremhæver, hvordan materialer i et pragmatisk perspektiv er mere end det fysiske stof som vi former en idé ud af. Snarere former og strukturerer materialer hele design processen, ved at hjælpe os med at indramme problemer, inspirere løsninger, og afprøve disse løsninger i praksis. Denne forståelse er af materialer er udviklet gennem flere af de vedlagte artikler, og bliver her testet og illustreret i en række eksperimentelle design cases.

# Table of contents

# 1. Introduction

Materials are part of design processes: we use and expend materials when creating sketches and prototypes, and we work with materials in the process of shaping the products that results from a design process. This dissertation proposes understanding materials and design reflection through the lens of pragmatism, and argues that what design materials do must be understood in a wide sense: they structure our design work by supporting reflection, collaboration and crafting processes. The dissertations aims at answering my research question:

*How may we conceptualise the role of materials in interaction design processes?*

My research aims to develop an improved understanding of how materials support design processes, and in doing so, explores and combines two strands of the interaction design field: the evolving interest in materials and materiality in interaction, and participation in design processes. I propose that we understand design materials in interaction design processes through the lens of pragmatism, specifically philosopher John Dewey's concept of *technology*.

Design processes invariably take place in complex situations where any act has intended and unforeseen consequences, and many participants act at the same time and with different backgrounds and motives. This means that the questions and dilemmas explored while designing cannot be definitely settled: design problems are 'wicked' (Rittel and Webber 1973), meaning they cannot be definitively and exhaustively predefined – we must explore and "set" our problems throughout a design process (Schön 1983). To explore and understand how to conceptualize materials in design processes, I have been part of a number of research-through-design projects throughout the course of my PhD work, and explored aspects of participation by delving into contemporary Participatory Design (PD) research. All the cases in the PhD project are concerned with either design processes, participation or materials, and have been conducted using a research-through-design approach. Much of this work has been guided and shaped by the work of pragmatist philosopher John Dewey, which I have developed into a focus on understanding design materials as *technology*, a term which for Dewey is broader than everyday use. It is those parts of a situation that can be used to experimentally resolve a doubt, problem or curiousness (see chapter 5). By employing and developing this lens, this thesis presents a theoretical framework for understanding how materials support design inquiry. I include a set of cases that demonstrate the usefulness and purpose of a pragmatic perspective.

# Motivation and research areas

My work reflects a growing interest in materials and materiality in Interaction Design (IxD). My main concern during the last three years has been to investigate how we might use materials, digital and physical, to stimulate ideas and move design inquiry forward. Further motivation for my work has been provided by HCI and IxD research's growing interest in materials and materiality. This interest may be ascribed to the many new technological innovations in materials and materiality, which means that we are now able to go beyond the screen and input devices, and begin to combine what Ishii and Ullmer (Ishii and Ullmer 1997) labelled 'tangible bits' into composites of computation and materials (Vallgårda and Redström 2007; Vallgårda 2013). This means that research interests within the field have been increasingly directed at the materiality of the computer and the quality of the interactions with it (Robles and Wiberg 2011; Wiberg 2015). Although in 1990 Grudin (Grudin 1990) alerted us to how the computer 'reached out' into new foci on collaboration mediated by computers, the trend of focusing on the materiality of computation constitutes a return to the roots of our field, by focusing on new types and shapes of computers, thus setting the stage for new experiences. The interest in materiality in HCI has stimulated my work within design processes in two ways. First, I am interested in how materials frame multidisciplinary design processes, which is why this overview article also includes reflections on sketches, prototypes and external representations in general. Second, I am interested in how we, as designers, work with a diverse set of materials – a cursory glance at my publications will reveal that they deal with interactive architecture, tangible interaction and hacking old computers.

There is a potential and a need for developing conceptual and methodological tools to address the roles materials play in design processes. While in recent years the field has developed new ways of understanding materials and computation (e.g. (Ishii and Ullmer 1997; Vallgårda and Redström 2007; Bdeir 2009; Vallgårda and Sokoler 2010; Bergström et al. 2010; Robles and Wiberg 2010; Sundström et al. 2011; Telhan 2011; Jung and Stolterman 2011b; Robles and Wiberg 2011; Jung and Stolterman 2011a, 2012; Blevis, Lim and Stolterman 2006), there is still a need to develop theory that connects designerly ways of thinking (Cross 2006; Lawson 2006; Brown 2008) to the focus on materials as part of design processes. The work discussed is influenced by these two research themes – design processes and materials. This interest has been embedded in an environment and *experimental system* (Rheinberger 1997, 2010; Dalsgaard 2016) strongly influenced by Participatory Design. PD has a longstanding interest in creating situations of mutual learning (Halskov and Hansen 2015; Muller and Druin 2012; Simonsen and Robertson 2012) in multidisciplinary projects, and part of the motivation for my PhD work has been to unravel how we may understand participation, and how we might use materials to facilitate collaborative efforts.

My interest in a pragmatist perspective on materials began during the last year of my master's programme. On one occasion I was trying to make sense of a particular design experiment, a video recording of an Inspiration Card Workshop (Halskov and Dalsgård 2006) in which a multidisciplinary group of participants were generating ideas for a new product. I was familiar with Schön's example of Qvist, a teacher, demonstrating to Petra, a student at an architectural programme, how she might develop an idea by sketching to explore and solve how to let building and site fit together (Bamberger and Schön 1983; Schön 1992). During my exploration of using this theoretical approach to analyse my own video example, I was struck by how Schön's well-known example of design reflection cast materials as rather passive. The pen and paper used for sketching in Schön's example played a role as an external representation, yet the example was vague on the importance of the specific materiality of the sketching tools. What difference did it make, that what was available in the situation was pen and paper, as opposed to other materials? Might the same design reflection have been accomplished used post-its, CAD-drawing or something else? And what role did the skill in transforming the materials make to Petra's and Qvist's different approaches to problem solving? Is Qvist better at drawing than Petra, and does that make a difference? Might they have been equally good at reframing the question at hand, had other materials been used, which somehow leveled the playing field?

I did not manage to answer the above-mentioned questions during my master's studies, though my interest lingered, and forms part of my personal motivation to delve deeper into the pragmatist underpinnings of Schön's ideas, by examining Dewey's work. In the papers appended to this overview article, I have tackled this question and related questions of participatory design as a collaborative design process of mutual learning. Pragmatism offers three advantages in terms of my work here: 1) the focus on emergence and situations of doubt and resistance means it is well-suited to grasping design processes dealing with wicked problems; 2) casting materials as *technology* in a pragmatist sense gives us a way of conceptualising the transformation of both design ideas and design materials; 3) the potential for discussing participatory design as situations of learning, since pragmatism focuses on development and learning.

To explore my research question, I have carried out practical work in the Creativity in Blended Interaction Spaces (CIBIS) project [1]. CIBIS is a multidisciplinary project aimed at developing and exploring how we might blend digital and physical environments in ways that supports the creative potential of users. Furthermore I am part of the strategic and interdisciplinary research centre for Participatory Information Technology (PIT)[2]. The PIT centre aims re-conceptualizing participation, while providing alternative ways of understanding, developing and deploying IT.

---

[1] http://cavi.au.dk/research-areas/cibis-creativity-in-blended-interaction-spaces/

This dual focus on design processes and my background in Participatory Design is reflected in my work. I have carried out surveys of related work and extensive literature reviews, one of which was published, the others forming the basis for papers published on other subjects. My practical work has been involvement in design processes, either directly as in (Hansen and Dalsgaard 2012; Korsgaard et al. 2012; Hansen, Nørgård and Halskov 2014), while others e.g. (Hansen and Halskov 2014; Andersen et al. 2015; Halskov and Hansen 2015; Hansen and Halskov 2016) are either collaborations with others doing practice-based work, or literature reviews. Doing practice-based work is key, as it is difficult to explore and analyse design processes without taking part in them. The constant back and forth of design judgments (Nelson and Stolterman 2003) is difficult to document, let alone grasp, from the outside. Without direct involvement by either a co-author or me, these papers would have been impossible.

## Overview of the dissertation

This dissertation consists of seven separate chapters, including this first one, which includes an overview of both the overview article and the published papers. In chapter 2 I discuss my methodological considerations, in chapter 3 design processes, in chapter 4 design materials, and in chapter 5 pragmatism. After that, I discuss my contributions in chapter 6, before concluding with chapter 7. Here I first summarise each paper's main contribution before summarising how each of the papers discusses *design processes,* and to varying degrees, *participatory design*, *design materials* and *pragmatism* (see table 1 below).

---

[2] http://pit.au.dk/

| Paper title | Main contributions |
|---|---|
| The Productive Role of Material Design Artefacts in Participatory Design Events | Identification of how physical materials play a key role in Participatory Design events<br>Pragmatist framework for understanding physical design materials<br>Five design considerations for using physical design materials in PD |
| Odenplan – a media façade design process | Identifies the need for, and develops novel techniques for tackling media architecture design challenges |
| Crafting code at the demo-scene. | Framework for understanding coding as crafting<br>Identifies role of digital materials in coder practice |
| Material Interactions with Tangible Tabletops: A pragmatist perspective | Framework for analysing interaction with tangible tabletops<br>Development of pragmatist framework to encompass interactive systems<br>Conceptualises materials as pragmatic technology |
| The Diversity of Participatory Design Research Practice at PDC 2002–2012 | Fundamental aspects of PD from classic Participatory Design literature<br>Five main categories of research contribution<br>Identifies how participation is defined<br>Identifies how participation is practised |
| Participation as a matter of concern in participatory design | Highlights participation as a matter of concern rather than a matter of fact<br>Identifies three challenges for PD:<br>(1) participants are network configurations,<br>(2) participation is an aspect of all project activities<br>(3) there is no gold standard for participation. |
| Materials in interaction design processes | Develops pragmatist framework for understanding materials in design processes<br>Identifies three roles for materials: How do design materials 1) establish problematic situations, 2) help resolve them through inquiring strategies and 3) support transformation of the problematic situation? |

Table 1 – an overview of the papers and their contributions

## Paper 1: The Productive Role of Material Design Artefacts in Participatory Design Events

Nicolai Brodersen Hansen and Peter Dalsgaard. 2012. The productive role of material design artefacts in participatory design events. In Proceedings of the 7th Nordic Conference on Human-Computer Interaction: Making Sense Through Design (NordiCHI '12). ACM, New York, NY, USA, 665–674.

This paper discusses how physical design materials support participatory design events. Together with my co-author, I argue that analogue materials play a key role in many PD methods, yet there are few comprehensive discussions of this role. This paper contributes to *Participatory Design* and *design processes* by investigating how design concepts emerge and evolve, arguing that we must adopt a more systemic perspective, and study the interplay between human participants and the manifest resources in a given situation. To investigate this question, the paper uses an experimental workshop called 'The Living Blueprint', which engages librarians of a projected library in generating ideas for interactive installations. Based on this, we suggest a theoretical framework based on Deweyan pragmatism, focusing on the times in a design event that materials had a *productive role*, that is, helped move design inquiry forward. On this basis, the paper suggests five design considerations: *enabling rapid transformations*, *documenting decisions*, *aligning collaborative efforts*, *provoking reflection*, and *proposing and supporting design changes.* These five considerations may be considered additions to a reader's *repertoire* (Schön 1983), or the products of a detailed qualitative study of design practice.

## *Paper 2: Odenplan – a media façade design process*

Henrik Korsgaard, Nicolai Brodersen Hansen, Ditte Basballe, Peter Dalsgaard and Kim Halskov. 2012. Odenplan: a media façade design process. In Proceedings of the 4th Media Architecture Biennale Conference: Participation (MAB '12). ACM, New York, NY, USA, 23–32.

This paper, published at the *Media Architecture Biennale 2013*, contributes to the research area of *design processes* by taking the challenges of designing for media architecture as its point of departure. In the paper we analyse a series of design experiments conducted as part of an experimental research project that developed concepts for the projected Odenplan metro station in Stockholm, Sweden. By designing a workshop customised to tackle inherent design issues of designing for media architecture, the project both contributed to developing design processes fit for media architecture, and offered broader insights into design processes, highlighting novel ways of sketching on 3D models, which blended digital and physical materials. This paper addresses *materials*, in that all the media architecture challenges tackled are framed by, and developed through a process of using materials to imagine their potential for design.

## *Paper 3: Crafting code at the demo-scene*

Nicolai Brodersen Hansen, Rikke Toft Nørgård and Kim Halskov. 2014. Crafting code at the demo-scene. In Proceedings of the 2014 conference on Designing interactive systems (DIS '14). ACM, New York, NY, USA, 35–38.

This paper, published at the *Designing Interactive Systems* conference 2014, introduces the idea of craftsmanship as a way of understanding the shaping and reshaping of code. It does so by presenting a brief case study of a programmer creating digital art on an old computer. In following his work process, this paper argues that his practice of coding might be considered a form of crafting (echoing (Lindell 2013)), and that code may be seen as a material. The paper contributes to the discussion of design materials and design processes by showing how, in light of philosopher Richard Sennett's work (Sennett 2008), we might consider the digital as a material from a crafting perspective. Furthermore, Sennett's philosophy resonates with pragmatism in its focus on crafting as a constant conversation with materials.

## Paper 4: Material Interactions with Tangible Tabletops: a Pragmatist Perspective

Nicolai Brodersen Hansen, Kim Halskov. 2014. Material Interactions with Tangible Tabletops: a Pragmatist Perspective. In Proceedings of the 8th Nordic Conference on Human-Computer Interaction: Fun, Fast, Foundational (NordiCHI '14). ACM, New York, NY, USA.

In this paper, we investigate how interaction with tangible interactive tabletops may be seen as a material exploration of form and sound. The paper takes as its point of departure the 'Radar Table', an interactive tabletop that allows users to create soundscapes by manipulating tangible objects. The Radar Table was developed over the years at CAVI, our research lab. The paper explicitly addresses *pragmatism* by using theoretical concepts drawn from John Dewey's work to develop a framework that allows us to analyse interaction with the Radar Table when it is deployed 'in the wild', in this case, at a major Danish music festival. This paper served to develop my theoretical understanding of pragmatism, while also offering a coherent framework for analysing systems in use. Lastly, the paper shows how the *design materials* of the Radar Table might be conceptualised as *pragmatic technology* – to this end, the paper has a strong focus on how the material form of the installation shaped interaction, and was adopted by users.

## Paper 5: The Diversity of Participatory Design Research Practice at PDC 2002–2012

Halskov, Kim and Nicolai Brodersen Hansen. 'The diversity of participatory design research practice at PDC 2002–2012'. International Journal of Human-Computer Studies 74 (2015): 81–92.

This paper, published in the *International Journal of Human-Computer Studies*, is a literature review of a decade of research in the field of Participatory Design, that draws on a decade of full papers published at the Participatory Design conference (PDC). As the most esteemed PD conference, we felt that the PDC constituted a good point of departure for this discussion. Furthermore, contemporary discussions of Participation in the HCI field (e.g. at CHI) prompted us to take stock of what was already there, offering possibilities for positioning, and discussions related to the PD field. To that end, we focused on a) the contributions made by researchers and b) how they defined and conducted 'participation'. The paper offers four main contributions: it identifies five fundamental aspects of PD from key participatory design literature; it identifies five main categories of research contributions; it identifies how PDC researchers define participation; it identifies how participation is conducted by researchers in experimental design cases.

This paper contributes to my focus on *Participatory Design* by highlighting the diversity of the research environment of which I have been a part, through the strategic research centre for Participatory Information Technology. It also focuses on *design processes* by highlighting the interplay of PD and experimental work, thus seeding the ground for future work in this area, for instance, my paper 'Participation as a matter of concern in participatory design' which focuses more explicitly on my own experimental work.

## *Paper 6: Participation as a matter of concern in participatory design*

Andersen, L. B., Danholt, P., Halskov, K., Hansen, N. B. & Lauritsen, P. (2015). Participation as a matter of concern in participatory design. *CoDesign*, *11*(3–4), 250–261.

This paper was published in a special issue on intersections of Co-Design and Actor-Network theory, in the *International Journal of CoCreation in Design and the Arts*. It builds on the same study of PDC research on which we drew in 'The Diversity of Participatory Design Research Practice at PDC 2002–2012', which is used to highlight the paradox that while 'participation' is the key concept of participatory design (PD), in the PD literature there are few explicit discussions of what constitutes participation. This paper was a collaboration with the STS research group at our department, who are currently part of the Teledialogue project which aims to design an IT-enabled platform for communication between social workers and children placed in foster care or at institutions, through participatory methods such as design workshops and qualitative interviews. Taking its point of departure in Actor-Network Theory, the paper questions and develops the idea of participants as networks spread across reports, government institutions, boyfriends and social workers. The argument is synthesised as three challenges for PD: (1) participants are network configurations, (2) participation is an aspect of all project activities and (3) there is no gold standard for participation.

This paper is a development of our study of PD's history, and contributes to *Participatory Design* by questioning and developing a deeper understanding of a key concept in PD. The paper also contributes to *design processes* by highlighting how, throughout the design process, participants are drawn into the process, and leave it again at other points.

## *Paper 7: Materials in interaction design processes*

Hansen, Nicolai Brodersen, Halskov, Kim. 'Materials in interaction design processes' to be submitted at Design Studies

This paper, currently under review, examines materials as critical components of *design processes.* The main argument is that while materials play a key role in exploring design spaces and creating design solutions, there are few comprehensive discussions addressing how exactly this exploration of design spaces unfolds. In this paper we develop Dewey's pragmatist philosophy by framing materials as pragmatic technology and introducing and developing the concepts of situation, inquiry, and transformation into a coherent framework usable for understanding the role of materials in creative design processes. As the principal case, we investigate the design process of a media façade for the 2010 World Expo in Shanghai. Through this analysis we identify three important questions related to the roles of materials in design: how do design materials 1) establish problematic situations, 2) help resolve them through inquiring strategies, and 3) support the transformation of the problematic situation? This pragmatist framework constitutes the main contribution, and enables researchers to reflect on the role materials play in creative design processes.

In table 2 I have illustrated how my seven papers contribute to the four research areas of materials, design processes, participatory design and pragmatism.

| Paper title | Materials | Design Processes | Participatory Design | Pragmatism |
|---|---|---|---|---|
| The Productive Role of Material Design Artefacts in Participatory Design Events | X | X | X | X |
| Odenplan – a media façade design process | | X | | |
| Crafting code at the demo-scene. | X | X | | |
| Material Interactions with Tangible Tabletops: A pragmatist perspective | X | | | X |
| The Diversity of Participatory Design Research Practice at PDC 2002–2012 | | X | X | |
| Participation as a matter of concern in participatory design | | X | X | |
| Materials in interaction design processes | X | X | | X |

Table 2 – illustrating the research areas addressed by each paper.

# 2. Research approach and activities

This chapter presents and discusses my research activities and the environment in which these have been carried out. First, I introduce the overall research framework in which I have carried out my research. Then I introduce research-through-design as the approach to my work. The aim of this chapter is to discuss the merits and limitations of doing practice-based research when studying design processes, and describe how any doctoral research project is shaped, and in turn helps to shape, the environment of which it is part.

My work has been theoretically based on Interaction Design and HCI in which I have specifically focused on using pragmatism as a theoretical frame for understanding materials. The value of importing, developing and testing theoretical concepts has been discussed by, among many others, Rogers (Rogers 2012) and Stolterman (Stolterman 2008). My main context of work has been the Interaction Design and HCI-research group at Aarhus University. This group has historically focused on both design processes and Participatory Design, work that has been carried out both by individual researchers, as well as under various research grants with more specific aims. I have been lucky enough to be part of two of these larger research projects. My PhD is formally funded by the Creativity In Blended Interaction Spaces[3] research project. I am also part of the centre for Participatory Information Technology[4]. Furthermore much of my work is conducted using the lab facilities and technical resources of the Centre for Advanced Visualisation and Interaction (CAVI) (Halskov 2011). Last, I have completed a four month research stay at the School of Informatics and Computing at Indiana University, Bloomington, USA. There I visited professor Erik Stolterman, and the visit in its entirety proved a useful challenge to my positions as well as encouraged me to delve deeper into Dewey's philosophy. As I shall return to in succeeding sections, the environment at Aarhus University constitutes the *experimental system (Rheinberger 1997, 2010)* that my works draws on as well as contributes to.

## 2.1 Research through design

Christopher Frayling's influential article (Frayling 1993) on research in art and design have provided the point of departure for most reflections on research-through-design, with its definition of three kinds of research: research into art and design; research for art and design; and research through art and design. In Interaction Design and HCI, a host of contributions have developed the idea of research through practice into a solid body of literature on research-through-design (e.g. Archer 1995; Zimmerman, Forlizzi and Evenson 2007; Stolterman and Wiberg 2010; Basballe and Halskov 2012; Bowers 2012; Bayazit 2004). These contributions are brought together by a strong focus on using practical, 'in the wild' work to develop questions that could not be asked in the laboratory under more controlled circumstances. Zimmerman et al. (Zimmerman, Forlizzi and Evenson 2007) provide a set of examples, and key to research-through-design is a focus on using an experimental and exploratory practice. Stolterman (Stolterman 2008) discusses how such work has immense value because of the complexity faced by designers, which is a different sort of complexity than that faced by natural scientists: designers deal with *messy situations* (Schön 1983, 1987) or dilemmas, because design practice ultimately deals with *wicked problems* (Rittel and Webber 1973). To deal with this different kind of complexity, Stolterman (Stolterman 2008) advocates abandoning any idea of modelling design research on the natural sciences, since the two sorts of complexity faced are very different; any such attempt would be futile.

---

[3] http://cavi.au.dk/research-areas/cibis-creativity-in-blended-interaction-spaces/

[4] http://www.pit.au.dk

Instead the research community should be aiming at *theoretical grounding; the study of practice; rationality resonance; forms of design support*, and *interaction design research measure of success* (Stolterman 2008). Of these contributions, my dissertation reflects a desire to use and develop theoretical grounding for understanding the practice studied. The overall aim of the CIBIS project is, among others, to develop forms of design support that resonates with the rationality of practitioners. In some ways, my PhD work contributes to those aims too, because what I contribute is a theoretically grounded understanding of practice. I argue that such an understanding is vital to develop forms of design support, since theories are often ways of articulating the practice such design support systems aims at. This aim, while not specifically in focus for the papers, presented here, resonates with that of Stolterman and Wiberg (Stolterman and Wiberg 2010), who argues that *concept-driven* interaction design research is a necessary next step for the field. While the CIBIS project has not yet reached its culmination, it is an expressed aim to manifest theoretical ideas in designed artefacts, to try out the validity of theoretical stances. Binder and Brandt (Brandt and Binder 2007) have articulated the relationship between experimental design work and research questions as a *question, program* and *experiment*. They describe how each research project is guided by an overall question, that through a series of programmes (such as for instance a design project) elicits specific experiments. It is crucial that there is a tension between the three layers: changes in one might affect the others: insights from a specific experiment might lead to changed programme. Dalsgaard (Dalsgaard 2009) has highlighted how many projects are conducted in collaboration with non-academic partners, meaning that the model advanced by Binder and Brandt (Brandt and Binder 2007) needs to be expanded to take into account the fact that these partners also have aims, that might differ from an interest in academic questions. If, as I shall do in the succeeding sections, a PhD project is viewed as part of an *experimental system* it also makes sense to ask how this system influences the PhD. The tension between questions, programs and experiments are not just the only tensions to be faced as a researcher, we are also faced with the tension over overlapping questions.

Basballe and Halskov (Basballe and Halskov 2012) argues that while the field interaction design and HCI contains a host of reflections on research through design through artefacts, there are few discussions of the design process of these artefacts. I agree with this position, and the quite detailed articulations of practical design work in my papers reflect this interest. Such a position further motivates the development of theoretical frameworks, since such frameworks can be utilized to articulate what happens in design processes. A good example of the value of theory is provided by interest in *critical design* (see e.g. (Bardzell and Bardzell 2013, 2015), where the importing and development of critical theory for discussing objects of design have provided the field of HCI with a solid body of theory to draw on and move forward.

In summary my research approach has been Research through Design. Methodologically I have, apart from that, conducted literature reviews as well as participated in a host of other PhD activities such as seminars, workshops and presentations. While these did not yield concrete papers, they still contributed greatly to my scholarly development. Indeed many of the papers featured in this dissertation is the result of discussions and position papers at workshops at for instance DIS and NordiCHI.

## 2.2 Experimental systems and Research through Design

My project has been carried out as part of the environment at Aarhus University, which brings along with it a host of already embedded practices and research interests. It is a key point that no PhD project, or research project in general, is an island. Rather it might be considered an *experimental system* defined as the "basic unit of experimental activity combining local, technical, instrumental, institutional, social, and epistemic aspects." (Rheinberger 1997). Dalsgaard has developed the idea of experimental systems to a perspective on research-through-design (Dalsgaard 2016), which allows us to go some way towards accounting for the situatedness of design research projects. In defining *experimental systems* in research through design, Dalsgaard highlights how Rheinberger's conceptualisation of experimental systems breaks with the idea of science as a series of controlled experiments. Rather it is a vehicle for materialising questions, something that resonates well with design research as focusing on wicked problems (Dalsgaard 2016). An experimental system contains a host of things, but of special interest to this dissertation is the interplay of *epistemic things (*those phenomena and objects not yet fully known or understood), *technical things* (those phenomena and objects we do know and draw upon in our everyday work) and *graphemes* (representations and material traces of knowledge that are actively created). With time, epistemic things might become technical things – having understood and developed them we might incorporate them into our scientific practice. Going back to my research environment, we might say that the experimental system of the research unit here at Aarhus University encompass the CIBIS and PIT projects, CAVI and the Interaction Design and HCI group. Inside of this experimental system is a series of epistemic things of specific interest to this PhD project: the research interests in design processes and Participatory Design, of which the question of materials in interaction design processes forms a small part. In that way I, as a PhD student have entered into an experimental system, and naturally the projects reported on in each paper reflects the fact that the questions the papers answer were generated by this system. This accounts for the fact that media facades are featured in two of the papers appended (Korsgaard et al. 2012; Hansen and Halskov 2016) – the rest of the experimental system here has a keen interest in this subject. Rather than being a problematic influence, it is a precondition of doing experimental work in a larger research unit, and one that is often neglected to report on. In an experimental systems perspective, being a PhD student, or any kind of researcher is not having a separate project that runs on its own, in splendid isolation. Rather it means bringing a set of epistemic and technical things into the existing experimental system, and working with them. Going over my seven appended papers, they might all be considered *graphemes* in that they are representations of knowledge generated by the experimental system I participated in. Furthermore, many of the papers draw on graphemes – in most of the cases examined we drew on a system for design process documentation, the PRT (Dalsgaard and Halskov 2012) to document and reflect on the process as it unfolded. That too is an example of a resource already available in the experimental system I entered into at the start of my PhD. At the same time my work in using this system has lead to new inquiries into how we might document and reflect on design processes, showing how my participation reflects back on the experimental system.

In summary my work has entailed both elements of research through design as well as the methods employed in individual papers: hosting experimental participatory design workshops (Hansen and Dalsgaard 2012); participating directly as a designer (Korsgaard et al. 2012; Hansen and Halskov 2016); studying video material of design in use (Hansen and Halskov 2014); participant observation (Hansen, Nørgård and Halskov 2014), and literature reviews (Halskov and Hansen 2015). These approaches all have advantages, but obviously also limitations. First, I am deeply embedded in all the practical design work, and all but the literature reviews deals with non-replicable studies. This, while a precondition of contemporary design research, still means that my project and collaborators have an extra obligation to establish and convince our peers of the rigour and accountability of our work. Frauenberger et al. (Frauenberger et al. 2015) have, albeit in a PD context, argued that rigour and accountability are nuanced concepts that are delivered through debate and argumentation. While the papers appended at the end of this overview article attempts to establish such rigour and accountability on their own, I have also attempted to describe how the overall direction of my work has been shaped by the experimental system I have participated in.

# 3. Design Processes

In this chapter I advance three points about design processes and participatory design: First that design processes concerns the experimental and iterative working through wicked problems. Second that design methods adds to a designer's repertoire by crystallizing the experiences of design researchers into a form that a designer can appropriate in unique design situations. And third that such appropriation depends on the ability for on-the-spot reflection of the designer. Furthermore I discuss participatory design as the background for, as well as the context within which my work has been conducted.

Interaction design is a design discipline in which designers deal with *wicked problems* (Rittel and Webber 1973) through a learning process of experimental moves (Schön 1983, 1987). These experimental moves are never purely intellectual exercises, but rather involve materials of all kinds. Wicked problems are characterised by having no definitive formulation, no stopping rule, and right or wrong solution (Rittel and Webber 1973). Design processes deal with *wicked problems* because design processes concern dilemmas that are at the same time *ultimate particulars* (Stolterman 2008; Nelson and Stolterman 2014). What constitutes a design problem is negotiated throughout the design process (Buchanan 1992). Design processes unfold through a constant exploration of the *design situation* (Löwgren and Stolterman 2007; Nelson and Stolterman 2014), which is both the reason for, as well as the context in which a design process takes place. A design process is thus aimed at transforming a given design situation from its initial point of departure to its completion; at exploring, defining and transforming the wicked problem at hand. Such a definition is never final, but is rather a temporary *framing* (Schön 1983), that will be tested again and again throughout the design process. This means that choices in design processes are not strictly rational and logical, but rather relies on a constant process of design *judgment* (Nelson and Stolterman 2003). Such judgment relies on the designer's background and repertoire, on what has been called abductive thinking (Kolko 2010).

Design processes, while all unique, still share some typical challenges, and within research literature, such challenges are met by developing design methods or techniques. One typical example might be how to gather knowledge about a new domain through field studies, how go generate ideas from sources of inspiration (Eckert and Stacey 2000; Halskov and Dalsgaard 2008; Halskov 2010; Kwiatkowska, Szóstek and Lamas 2014), how to test ideas through sketches (Schon and Wiggins 1992; Purcell and Gero 1998) or prototypes (Floyd 1984; Lim, Stolterman and Tenenberg 2008; Hartmann 2009). In all of these methods materials play a key role – they are part of framing the exploration of the constraints and possibilities of a design. Thus, from the perspective of this PhD project, materials play a key role in design processes by offering up ways of exploring and transforming designs.

## 3.1 Participatory Design

Participatory Design (PD) has played an influential role in my work. Both as part of the context I work in, as a specific research interest. I first outline the background for PD, focusing on democracy and mutual learning, methods and materials as part of methods.

Participatory Design evolved out of a specific Nordic model for cooperation among workers and employers, the main objective being improving knowledge of systems in context; establishing realistic expectations and reducing resistance to change; and increasing workplace democracy by giving members of an organisation a voice in the design process (Bjerknes and Bratteteig 1995). While the two first reasons can be considered practical and drawing on an *integrationist* perspective (Muller and Druin 2012), the latter is political*,* and reflects what can Muller and Druin refers to as a *conflict* perspective. Seminal PD literature works (eg. Schuler and Namioka 1993; Greenbaum and Kyng 1991) stress methods as an important aspect of PD as a democratic endeavour, aimed at creating situations of *mutual learning,* using for instance prototypes to enable users to have a voice in the design process (Simonsen and Robertson 2012). Iversen et al. (Iversen, Halskov and Leong 2010) consider values as main driver in PD, highlighting the need for discovering them, developing them and grounding them in the final design. This perspective takes a view of PD as somewhat agnostic of specific method, in that it is not so much the specific approach, but rather the stance towards designing with people that matters. As a contrast to this, "Participatory Design" is sometimes used in a less political and value-driven meaning of the word, like for instance Sanders et al. (Sanders, Brandt and Binder 2010) who, following the *integrationist* agenda discussed by Muller and Druin (Muller and Druin 2012), offers up a framework for organising the tools and methods of PD. So while other development methodologies such as User-Centered Design (UCD) recognize the value of learning from users, PD places a special emphasis on allowing users to be part of the decision-making, and being part of generating alternatives. As part of my work we have reviewed the most prominent conference within the field, the Participatory Design conference (PDC) (Halskov and Hansen 2015). Through this literature review we emphasise how *politics, people, context, methods* and *products* in different combinations according to research interests can be said to constitute the core of modern PD research, though often methods are used to bring one of the other aspects into play. Therefore we might say that mutual learning through methods becomes a pathway to the democratic participation of users in design processes.

PD methods help establish situations of democratic dialogue and mutual learning by enabling users to have a voice in the design process through for instance prototypes, (Simonsen and Robertson 2012). Defining methods as ways of creating mutual learning aligns well with an integrationist perspective on PD (Muller and Druin 2012) where the aim is integrating insights from different disciplines. And drawing on Greenbaum and Madsen (Greenbaum and Madsen 1993) one might say that mutual learning is a pragmatic aim. That is not to say that a political- or conflict-perspective (Muller and Druin 2012) is incompatible with using methods to create situations of mutual learning. Rather, establishing situations of mutual learning is a precondition for creating democratic design processes. Mutual learning is a process of different disciplines coming together to inquire into a design question. Therefore mutual learning is a key factor and prerequisite for both conflict- and integrationist forms of participatory design and it makes sense to ask how these situations of mutual learning develop, and one aspect of this is materials as part of design processes.

However, like other design methods (discussed by ie. (Stolterman 2008)), PD methods can only serve as a starting point that must then be appropriated to the specific and particular circumstances of the project. In her PhD dissertation, Eriksen has highlighted how materials assist in "formatting" PD activities (Eriksen 2012), meaning that a method is highly intertwined with its material components, which are then negotiated and changed throughout a design activity. This resonates with Light and Akama (Light and Akama 2012) who argue that "participation" is more than just a method, and critiques a narrow focus on design methods as described in academic research. According to Light and Akama much depends on the "enactment" of the method, and the necessity of reflecting on the role and agency of the practitioner in relation to these methods. Thus, while PD does have a set of established methods, familiar to most researchers and practitioners in the field, the application and staging of participation is a core issue still to be tackled. Such reflections draw our attention to the need for articulating the reasons for choices in PD processes, something that has been described as the need for a language of reflecting and argumentation for stances, choices and judgements (Frauenberger et al. 2015). We have ourselves called for greater clarity in articulating the details and choices of PD processes (Halskov and Hansen 2015), and following Frauenberger et al. (Frauenberger et al. 2015) one might say that developing a language of how materials play a role in PD processes contributes to such articulation work.

# 4. Design materials

Materials play two roles in interaction design processes: materials are the matter from which we craft products, meaning that engaging with materials during the process is as important as in other design disciplines and crafts. This has given rise to an interest in computers as material things. And in design activities, materials are ubiquitous, parts of design methods such as sketching or prototyping: even the most basic brainstorming method may involve Post-its®, a white board or a notebook. This means that there is a longstanding interest in materials for design thinking in interaction design.

Below, I outline the interesting traits of *computers as material things*, and *materials for design thinking*. At the end of this chapter I discuss *crafting* as a specific view in which a focus on materials is brought together with a design process focus. I propose that viewing materials through a pragmatist lens might bring all three framings together.

## 4.1 Computers as material things

With the advent of new materials that combine computational technology with other materials interaction design and HCI have increasingly been interested in how we might understand computers and experience as material. Much of this work draws explicitly or implicitly on that of Manzini, who describes how our understanding of a material shapes our ideas of what we can do with it, and vice versa: new materials prompt new ideas for products and objects (Manzini and Cau 1989). Similarly, Doordan (Doordan 2003) describes how materials used in design offer a way to focus insights from different disciplinary perspectives and methodologies. From Doordan's perspective, materials are part of the design problem to be solved through a design process. He exemplifies this by discussing how plastic (an example also used by Manzini) by its very nature complicates any efforts to talk about it, because it may take so many forms, shapes and textures. Thus, there is a strong relationship between the materials available for designing, and what goes on in the design process. The availability of new materials that challenge the status quo of the computer as a mouse, keyboard and screen is what gave rise to a new focus on the materiality of the computer itself, as opposed to just focusing on what happens on the screen. This has been called 'the material turn' (Robles and Wiberg 2010) or 'the material move' (Fernaeus and Sundström 2012), reflecting the availability of a new and very diverse set of materials. Such an increased availability of materials may also be described as what Manzini (Manzini and Cau 1989) call a *hyperselection* of materials. In such circumstances (exemplified with plastics, by Manzini), materials are available in such a variety and combinability that it no longer makes sense to talk of a set number of materials. Instead, materials are made to order, or as Vallgårda and Redström would have it (Vallgårda and Redström 2007), crafted into *composites* that fit a given purpose. This specific radical shift from *materials* to *materiality* (Robles and Wiberg 2011) has given rise to new and exciting investigations of the computer as a material thing, which may and must be combined with other materials in new ways. Gross et al. (Gross, Bardzell and Bardzell 2013) sum up three current and competing views of materiality in HCI; Tangible User Interfaces (TUI) as physical materials; computation as material; and crafting understood as materiality communication tradition. The first two views may be said to be metaphysical, in that they ask and describe what materials are and do, while the last focuses on how our use and perception of materials are culturally and historically situated. Gross et al. then bring together these three views by importing and developing a theory of materials as a *medium*. Thus, Gross et al.'s main contribution is a theoretical lens through which we might explore materials from physical, metaphysical and communicative perspectives.

Ishii and Ullmer developed the idea of *tangible bits* in a paper for CHI 1997 (Ishii and Ullmer 1997). This paper has been very influential, and highlights the physicality and materiality of computation, by showing how computation might be used in conjunction with a range of physical objects and metaphors, thereby heralding a shift towards Tangible User Interfaces (Shaer and Hornecker 2010). Ishii et al. later developed this perspective into the idea of *radical atoms* that are 'beyond tangible interfaces by assuming a hypothetical generation of materials that can change form and appearance dynamically, so they are as reconfigurable as pixels on a screen' (Ishii et al. 2012). This powerful idea reflects an interest shared by Vallgårda and Redström (Vallgårda and Redström 2007) who consider computational technology a *composite material*, that is, a combination of two or more materials, either enhancing a specific property or giving rise to new properties. Arguing that computational technology has substance, Vallgårda and Redström (Vallgårda and Redström 2007) advance the field by offering a computer-science perspective on materials that draws our attention to how *temporal form* (Vallgårda et al. 2015) is one of the key qualities of computational objects. As part of this understanding of the new and complex interplay of materials and computation, Fernaus and Sundström (Fernaeus and Sundström 2012) highlight the need for material explorations during the design process (as opposed to materials being selected after their function has been determined. Fernaus and Sundström discuss the impact of what they call 'The Material Move' in design, caused by the burgeoning array of materials available to designers. This necessitates both richer descriptions of how and when materials matter in interacting design processes, and detailed examples of specific designs using new materials.

Industrial design has long recognised the experiential qualities of materials – it matters greatly what a product is made of. The advent of new, computationally enhanced materials and new designs using such materials, has given rise to a renewed interest in how materials shape experiences. Bergström et al. (Bergström et al. 2010) emphasise how the development of 'smart materials' enables a whole range of new experiences. Recognising that new materials have always challenged existing design practices and given rise to new products (Manzini and Cau 1989), Bergström et al. argue for the development of concepts and theories that can better our understanding of these new materials and the experiences they facilitate. This thread has been followed up by Robles and Wiberg (Robles and Wiberg 2011), who argue that we are increasingly shifting from new materials to new materialities, meaning new experiences and understandings of what materials are and may be. Vallgårda et al. (Vallgårda et al. 2015) tackle these questions, and argue that interaction design is distinguished from most other forms of design by its attention to temporal form – computation has a temporality, and so do the physical materials that computation affects. By shifting states, thus altering the form of a computational object thing, it is possible to create interactive experiences akin to music and poetry, in which tempo plays a key role. By beginning to develop a theory that highlights interactive experiences with computational things, Vallgårda et al. (Vallgårda et al. 2015) develop an understanding of interaction designs as material experiences. Karana et al. (Karana, Pedgley and Rognoli 2015) advance the concept of *materials experience*, referring specifically to Doordan's concept of the *appreciation* of materials (Doordan 2003). To Karana et al., focusing on people's experiences of materials both immediately and over time, it becomes clear that materials exhibit both *experiential* and *functional* qualities. This needs to be considered in the designers' *material selection* (Karana et al. 2015).

In summary, the research interest in the *material move* has focused on investigating and discussing the role of computers as material things that are combined with other materials, in order to create new materialities (Robles and Wiberg 2011). New materials gives rise to new experiences because the materiality of the interactions change (Wiberg 2015). This focus on new materialities, while tentatively explored in papers 4 (Hansen and Halskov 2014) and 7 (Hansen and Halskov 2016), has mostly been a precondition for my work on design processes in which I have explored how materials are used and consumed as part of a design process.

## 4.2 Materials for design thinking

During a design process, designers use materials other than those they intend to use in the final product. These materials may be consumed during the design process, such as sketches or inspiration cards (Halskov and Dalsgård 2006), or investigate one part of the final composition of a product, for example, prototypes. In this section I review the ways in which materials have been used for design thinking (Dorst and Cross 2001; Cross 2006; Lawson 2006; Brown 2008; Dalsgaard 2014). I discuss three related framings of materials for design thinking: sketches, prototypes and a wider theoretical perspective of "external representations". Whereas sketches and prototypes focus on specific artifacts, considering materials as external representations means drawing in a specific theoretical perspective. This perspective has been included because it has been widely influential in the field of interaction design.

### *Sketches and sketching*

Sketching and sketches may be said to be prototypical design materials, discussed in almost any influential book on design processes. Indeed, Buxton (Buxton 2007) considers sketching to be the pivotal design activity, that all designers have in common. Sketching refers to the process of drawing informal, open and unfinished marks-on-paper to gradually work through a design idea. This focus on the interplay of ideation and sketching is evident throughout the literature on sketching, as is a focus on using materials other than paper and pen.

#### Sketching and ideation

Concerning the interplay of ideation and sketching, Schön and Wiggins (Schon and Wiggins 1992) discuss how the process of architectural designing may be seen as experimental work that involves seeing-moving-seeing, that is, the experimental back and forth between doing and appreciating. The emphasis is on seeing: an architect uses his or her appreciative judgment to determine a potential next move. Schön and Wiggins are deeply rooted in pragmatism (e.g. see (Bamberger and Schön 1983; Schön 1987, 1992, 1983)), and point out that the a) any account of designing must take the medium into account, b) *move experiments* involve several kinds of seeing, c) discovery is part of drawing and d) designing serves as preparation for further designing, by building up examples from which we draw.

Through a meta-review of protocol studies, Purcell and Gero (Purcell and Gero 1998) examine how sketching develops ideas during the design process. They argue that research into the value of sketching is concerned with themes such as reinterpretation, properties of sketches, generating knowledge, sketching as a cyclic dialectical process, and lastly, that expertise plays an important role. They also point out some unresolved questions about sketching: are particular kinds of drawing better suited to bringing about the above advantages? Do expert designers sketch in certain ways? And are particular ways of sketching associated with higher quality outcomes? To begin to tackle some of these issues, Purcell and Gero suggest that we import theoretical frameworks and concepts from cognitive psychology, and argue for the value of concepts such as short-term memory, imagery reversal and creative synthesis.

Stolterman advances the view that *formative skills*, such as the ability to imagine, are key to designing (Stolterman 1999), and show how designing relies more on judgment than on rational knowledge. Sketching aids in such formative work mean that materials and skill must play a key role. Stolterman emphasises this through the concept of 'diathenic graphologue' (also covered in detail in (Nelson and Stolterman 2014)), the process of 'letting a thing be seen through its image'. In this way, sketching is brought together with the formative skills needed for design – sketching aids imagination.

Tholander et al. (Tholander et al. 2008) emphasise that in the act of sketching, sketches, thoughts and bodies are not separate entities. By examining sketches on a whiteboard, they emphasise how this process is a deeply intertwined activity in which embodiment, the physical act of drawing, played a key role. Tholander et al. stress that neither the sketches nor the talking, gesturing or thoughts of the designers can stand alone – neither of these aspects, each interesting on its own, captures what really goes on in the process of sketching. Taken out of the context of the designers' talk and action, the sketches provide only a limited account of the system being designed. By gesturing and pointing, the designers illustrate the imagined use of, and interaction with, the sketched system.

### Sketching with new materials

On the topic of materials other than pen and paper, Löwgren (Löwgren 2004) develops the idea of sketching to take into account digital materials too, by highlighting the need for design representations that explore the intended use situations in some detail, and appear tentative enough to afford participation by users and stakeholders. He does so by presenting 'animated sketches', small animated movies that express important scenarios of the intended use of some imagined artefact. The sketches, while tentative, still enables the designer to express ideas that would be difficult to express using prototyping techniques. As such, the animated sketches might be used as 'vehicles for thinking', facilitating participation and collaboration, and finally as a rhetorical artefact for convincing clients of the value of an idea.

In their broader discussion of *designerly tools*, Stolterman et al. (Stolterman and Pierce 2012) discuss sketches as one potential design tool, before they present an investigation into the 'design-tool relationship' conducted through a series of interviews with designers, in order to advance the understanding of how practising designers use, understand and interact with their tools. Specifically, they emphasise how such tools were used to generate new ideas during the design process. Here, sketching is just one of many potential tools, and what makes a tool suitable for designerly work, is its ability to be incorporated into the individual designer's practice.

Brynskov et al. (Brynskov, Lunding and Vestergaard 2012) present the DUL radio, a way of sketching directly in hardware. The DUL radio is a small wireless toolkit for sketching sensor-based interaction. The reason for including it here is that the technology itself reveals a desire to sketch with hardware, and Brynskov's explicit motivation was to ease the process of sketching. This draws our attention to a) a recognition of the value of sketching and b) the desire to develop tools for sketching.

In summary, sketching as and approach to materials highlights how sketching is a process intertwined with thinking, emphasising the value of ambiguity/openness (to leave room for imagining) and rapid changes to the sketch (to allow the materials to follow the rapid mental developments when sketching). The materials for sketching may be paper-based or digital, but historically, sketching has been considered a physical process, since sketching is meant to be an informal and fast-paced activity, something that has historically meant using paper and pen. However, recent work has tackled sketching in other materials, such as video, or directly in hardware.

### *Prototypes*

Prototypes, typically employed later in the design process than sketches, are also a design material with which most designers and researchers will be familiar. When utilising prototypes we select certain qualities of a design idea, and try them out by creating a model of the idea. This model has to be of sufficient quality and functionality to address the qualities being explored, and the material qualities of prototypes interplay with the process of prototyping ideas.

Floyd (Floyd 1984) describes how prototypes are a contested term in software development and design, because unlike in industrial production, prototypes in design are experimental rather than specific tests of a finished idea. She describes how prototypes must be considered a vehicle for learning, and notes that a prototype will always have a scope. Floyd differentiates between horizontal and vertical prototyping, which refers to how many of the functions are implemented (horizontal) and whether these functions actually work (vertical). The choice of how the prototype should be set up hinges on the purpose of learning, the aspects and qualities of the design idea that the designer wishes to investigate.

The focus on learning is echoed by Mogensen (Mogensen 1992), who presents the concepts of *provotypes*, which are prototypes that provoke. Mogensen's argument is that there is a dialectical relationship between the desire to create new ideas, and making these ideas fit existing practices. This relationship may be foregrounded and explored through prototypes that seek to explore the 'why', 'what' and 'who' of practice, casting the designer as a provocateur, setting out to discover discrepancies. These discrepancies may then serve as openings for design, and shed new light on existing practices by forcing participants to do their current work in new ways.

Lim et al. (Lim, Stolterman and Tenenberg 2008) highlight how HCI and design seem to focus more on prototypes for *evaluation* than for *exploration*. To Lim et al. prototypes are 'a way of traversing the design space'. Lim et al. address prototypes as vehicles for exploration and discovery, and relate them to the idea of exploring a design space by filtering out certain objects. The incompleteness of prototypes lends them much of their power, by focusing on certain aspects of a design idea, and ignoring others.

The focus on exploration has been developed by Hartmann (Hartmann 2009), whose PhD work investigates prototyping as 'the fundamental activity that structures innovation in design'. To Hartmann, prototyping interactions for ubiquitous computing require a new set of tools, some of which he develops. He also highlights how prototypes are just a means to an end, echoing the authors mentioned above. This means that any sort of prototyping tool must support both exploration and iteration.

Sanders and Stappers (Sanders and Stappers 2014) consider prototypes as inhabiting a space between the generative and the evaluative phases of a design project. In doing so, they highlight how prototyping might be used both to produce ideas, insights and concepts that may then be designed and developed, and to assess, formatively or summatively, the effect or the effectiveness of products, spaces, systems or services.

In summary, prototypes are usually employed later in the design process than sketches, and the research reviewed here focuses much on how prototypes are used to test out already formed ideas.

## *External Representations*

The ideas of externalisations and representations have played significant roles in theories surrounding the role of the immediate environment in design processes. Thus, in HCI and Interaction Design, materials, prototypes and sketches are often conceptualised as external representations aiding design thinking (Cross 2006) and reflection-in-action (Bamberger and Schön 1983; Schön 1992; Schon and Wiggins 1992). External representations are a broader concept than design materials: words are a prime example of something that falls outside the scope of this dissertation, since I have chosen to focus on those physical and digital materials that are transformed, and consumed during the design process. A prime example of this is the sketches, cards, and prototypes we use.

### Representing and transforming design problems

In an early example of the idea of computer science as an experimental practice, Bocker et al.(Bocker, Fischer and Nieper 1986) highlight how the limits to our thinking are often equal to the limits of our imagination and visualisation capacities. Bocker et al. highlight the strong link between how a problem might be represented, and how and whether we can understand and solve it. Thus, a way forward for computer science as an experimental practice is develop better representations. This allows users in Bocker et al.'s case study' to 'play' with a set of graphical representations, experimenting with special cases, as well as experiencing a deeper engagement with the practice of programming. Building on the idea of representation as an instrument for solving problems, Fischer et al. (Fischer, Nakakoji and Ostwald 1995) emphasise how design artefacts, too, need to evolve during the development and solution of complex problems. Developing a system for representing problems and solutions, they highlight how design artefacts need to be both expressive (mutable) and associative (combinable), to support a useful representation and transformation of a design problem.

Eisentraut and Günther highlight how individual problem-solving styles and the use of representations are highly intertwined (Eisentraut and Günther 1997). They primarily examined sketches used during the design process, and highlight how both the style of problem solving and use of representations are significantly related to the kind of problem at hand. This means that there is an interplay among the expertise, the approach to problem solving, the available representations and means of sketching.

The desire to use and develop representations and externalisations that work with new design challenges is also investigated by Dow et al., who investigate the usefulness of representations and externalisations for visualising and working with ubiquitous computing concepts (Dow et al. 2006). Having done so, Dow et al. offer guidelines, revealing deficiencies and potential when applying designers' current typical toolset to the challenges offered by ubiquitous computing.

Dix and Gongorra (Dix and Gongora 2011) discuss how external representations are ubiquitous in design as a way of making the tacit explicit, allowing unreflective and embodied action to become the subject of reflection. Drawing on Hutchins (Hutchins 2000), Dix and Gongorra describe how externalisation involves the active shaping of the world as an intellectual resource, and present four roles for external representations: informational, formational, transformational and transcendental (Dix and Gongora 2011).

### Generating ideas

The work of Warr and O'Neill demonstrates how external representations may be used to facilitate shared understanding among participants in design processes (Warr and O'Neill 2007). Warrer and O'Neill developed a creativity support tool, the EDC, and discuss how the creation and development of boundary objects facilitates problem framing, idea generation and idea evaluation. The participants in this study could use the created boundary objects to allow ideas move from mental images to visual and tangible representations, open to critique. Thus, the EDC facilitated a shared understanding and detailing that went beyond what could be achieved through verbal communication.

Dittmar and Piehler (Dittmar and Piehler 2013) investigate how different design teams might use representations and externalisations (here in the form of QOC diagrams, HOPS models, and Java prototypes) to develop and connect a design space. They highlight how design ideas are gradually shaped by examining them through the 'lenses' of different internal representations throughout an experimental design process. By doing so, Dittmar and Piehler highlight how design ideas and representations co-develop and reflect each other.

The focus on idea development and material representations has also been investigated by Bjørndahl et al. (Bjørndahl et al. 2014). In their paper from 2014, Bjørndahl et al. present a taxonomy of the roles material presentations play in joint epistemic processes. According to Bjørndahl et al., material representations do their work for collective reasoning by supporting collaborators to illustrate, elaborate and explore new ideas.

### Facilitating participation

Coming from a Participatory Design approach, Kyng discuss how 'representational artifacts' gain their power from representing only a few select qualities of what they represent, their 'representational qualities' (Kyng 1995). In addition to this, they possess 'non-representational qualities', these being the qualities gained from a map that represents a city being made of paper. This interplay of representational and non-representational qualities allows participation to unfold both by representing only the qualities understandable and important to users, and by allowing users and designers to utilise non-representational qualities, for instance, those of a paper prototype.

Arias et al. investigate the collaborative aspects of using representations (Arias et al. 2000). Through a discussion of an experimental system, they highlight how we might tackle the challenges of collaboration among users with different backgrounds. Among other things, in this system, the EDC, representations and externalisations are used to establish a shared understanding among various stakeholders, contextualise information for the task at hand and create objects to help to think in collaborative design activities.

Bratteteig and Wagner (Bratteteig and Wagner 2012) compare three urban projects in which they conducted a series of participatory workshops. They discuss the interplay of representations, participants and the site, which generates and shapes design ideas, when using a novel, mixed-reality tabletop, the ColorTable. They both note that it is crucial that representations of the site be editable, as is the role of participants being able to imagine highly divergent and novel interventions at the three different urban sites.

In summary, considering design materials as external representations draw our attention to how they function as an intellectual resource in the world. By identifying the three roles from the research literature cited above, we see how an external cognition perspective might resonate with pragmatism which views activity as always taking place in an environment. However there is still a very clear distinction between what goes on in the mind and what happens in the representational artifacts.

## 4.3 Crafting

Contrasting the section on computers as material things with the section on materials for design thinking above, a fault line becomes obvious. It might seem that the first section refers to an engagement with the materials themselves, whereas the second refers to developing ideas, which are then subjected to a function-driven materials selection. The idea that we have a mental model and function that we then impose on a selected material, shaping the clay to reflect our imagination, is an example of what antropologist Tim Ingold calls the *hylomorphic* model of designing in his seminal book "Making" (Ingold 2013). In this way of thinking materials are bent to fit a particular mental image in the mind of the maker. However in Ingolds understanding, we do not move from materials to objects. Rather, there is a constant *flow* of both consciousness and of the (material) world, and what we choose to call a mental image is just one point in the flow of consciousness, and what we choose to call an object is just one point in the flow of materials. Thereby Ingold dissolves the distinction between the external world of materials and the internal world of ideas: "Far from standing aloof, imposing his designs on a world that is ready and waiting to receive them, the most he [the maker] can do is to intervene in worldly processes that are already going on, and which give rise to the forms of the living world that we see all around us – in plants and animals, in waves of water, snow and sand, in rocks and clouds – adding his own impetus to the forces and energies in play." (Ingold 2013 p. 21).

With the added focus on materials, HCI and Interaction Design have increasingly broadened their perspectives to encompass new ways of understanding. Of these, crafting deserves special mention since it specifically focuses on the interplay of designer and material, something discussed by both Mccullogh (McCullough 1998) and Gross et al. (Gross 2013; Gross, Bardzell and Bardzell 2013). Much of the work framing working with materials as craft draws on Sennett (Sennett 2008) and/or Pye (Pye 1968) e.g. (Tung 2012; Lindell 2013; Golsteijn et al. 2013; Zoran and Buechley 2013; Tsaknaki, Fernaeus and Schaub 2014; Baader and Bødker 2015; Rosner, Ikemiya and Regan 2015). What generally characterises much of the work in this category is that it focused on how designing might be seen as a conversation with materials as Sennett (Sennett 2008) and Schön (Schön 1992) would have it.

In his 2015 journal paper, Wiberg (Wiberg 2015) discusses how analogies to craft and crafting are recent developments due to the fact that computers are no longer only associated with transforming information, but also with physical materials. In Wiberg's words, researchers use the crafting analogy to turn a 'material lens' on interaction design, drawing inspiration from traditional crafting practicesy doing so, computational materials are, according to Wiberg, placed on equal footing with other materials in interaction design, aligning interaction design with other traditions of form-giving.

Tsaknaki et al. demonstrate such an integration of interaction design with traditional form-giving through material explorations, by exploring the potential of using leather and computational components (Tsaknaki, Fernaeus and Schaub 2014). In doing so, Tsaknaki et al. highlight the potential of using a 'classic' crafting material such as leather to create interactive experiences, and to discuss how new tools such as laser cutting and classic leather-working techniques may inform each other. Zoran and Buechley also discuss the convergence of information technology and craft, by discussing how the creative processes of digital fabrication and traditional crafting might be combined (Zoran and Buechley 2013). Their aim is to elucidate the resistance and 'workmanship of risk' (Pye 1968) of crafting, with the precision and replicability of digital fabrication. Golsteijn et al. (Golsteijn et al. 2013) present the idea of 'hybrid crafting', meaning crafting that includes both physical and digital components. By doing so, Golsteijn et al. highlight how we might imbue everyday creative practices of crafting with both physical and digital components, and discuss how the crafting process might use both physical or digital materials as the point of departure. The value of the work of is double – it both serves as a material exploration (Wiberg 2013) and establishes symmetry between physical and computational materials, meaning that both play equally important roles in design.

Last, it is worth mentioning the idea of crafting as a metaphor for coding – especially worth noting is the work of Lindell (Lindell 2013), who investigates how programmers frame their own practice. Lindell's is interested in the materiality of information technology, and through a survey of programmers found that 'material' and 'crafting' are useful metaphors for describing the practice of coding. Lindell describes how a description of crafting code resonates well with Sennett (Sennett 2008) and Schön (Schön 1983, 1992), a point echoed by Baader and Bødker (Baader and Bødker 2015). It is crucial to recognise that in recent literature both the word 'crafting' and the word 'material' may refer to a diverse range of practices and meanings. The main distinction is between physical materials and digital materials, since it is exactly this convergence of digital and physical practices that has given rise to the growing interest in crafting.

Having discussed materials as the stuff that information technology is created from, as used in design thinking, and as in crafting, it is now possible to begin to discuss how all three perspectives might be brought together through pragmatism.

# 5. Pragmatism and design materials

Pragmatism is a philosophical tradition, which has its roots in an empiricist attitude to the world: questions must be resolved through experimentation. It is a praxis-based philosophy in which categories and ideas of absolute truths are dispensed with in favor of a situated view of existence in which an idea, theory or object is evaluated based on its consequences in a specific situation. Such an attitude fits well with design processes as a continuous exploration of problem and solution. As such pragmatism offers a coherent and fully developed theoretical school of thought that others have already used to frame reflective thinking in design (e.g. (Östman 2005; Melles 2008; Dalsgaard 2014; Rylander 2012). In this chapter I first outline the aspects of Dewey's pragmatism I have chosen to focus on, before elaborating how we might consider design materials pragmatic technology, when considering how materials play a role in participatory design processes.

Many of the papers (1, 4, 7) produced during this PhD explicitly draws on the work of John Dewey to discuss material aspects of design processes. The rest reflect the empiricist attitude on a more general level, something that is a general trait of design processes. My work is based primarily on three books by Dewey: "Art as Experience"

(Dewey 1934), "Logic: The Theory of Inquiry" (Dewey 1938) and "How We Think" (Dewey 1910).

While "Art as Experience" from 1934 is considered Dewey's major work on aesthetics, it also contains an elaborate and coherent discussion of existence and experience as always being part of an ever-changing ecology: *"Life itself consists of phases in which the organism falls out of step with the march of the surrounding things and then recovers unison with it – either through effort or by some happy chance. And, in a growing life, the recovery is never mere return to a prior state, for it is enriched by the state of disparity and resistance through which it has successfully passed."* (Dewey 1934, p. 12-13) This is process is what Dewey refers to as "experience", and certain of these experiences stand out, and become problematic situations, worthy of inquiry. A few points must be stressed concerning Dewey's understanding of experience and existence. First we are always engaged with our environment – we do not just live in an environment, we live through and in exchange with the environment. Second the environment (which might consist of other people and physical things) is also changing – it is this "march of the surrounding things" that Dewey refers to in the above quote. Third, recovering balance with these surroundings must by necessity draw on them as well – Dewey makes no distinction between the internal and external, as also discussed by Garrison (Garrison 2009). And fourth, as we progress through the experience and reestablish the stability of the situation, we grow from it, and later become able to draw on that experience.

"Logic: The Theory of Inquiry" published in 1938, is Dewey's attempt at a systematic pragmatist answer to formal logic. As such the book is concerned with how logical propositions arise, are used and transformed. Logical propositions arise through *inquiry*, through action directed at specific situations: *"Inquiry is the controlled or directed transformation of an indeterminate situation into one that is so determinate in its constituent distinctions and relations as to convert the elements of the original situation into a unified whole"* (Dewey 1938, p. 108). The focus for my papers has been how inquiry unfolds and specifically what role materials played in these transformations of the problematic situation.

"How We Think" from 1910 is Dewey's work on how we might begin to separate good thinking from bad thinking, and how we might train ourselves for a rigorous and systematic approach to inquiry (Dewey 1910). Aimed at education, "How We Think" has been inspirational to my work because much design work can be said to utilize the kind of *systematic inference* of induction and deduction inherent in inquiring strategies.

By using pragmatism to frame my work on materials in participatory design processes, I gain two advantages. First, pragmatism is well-suited for the task at hand, with its focus on experimentation, emergence and no answers set in stone: to pragmatism any question might be considered a dilemma, in that each situation and problem is in essence unique. Second, because of the first point, by using pragmatism the PhD work has been in constant conversation with a well-developed school of thought within Interaction Design. Pragmatism has been influential in interaction design most notably through Schön, whose books (Schön 1983, 1987) has had a wide impact on conceptualizing the value of reflection in design processes (Baumer et al. 2014). Pragmatism shares the empiricist attitude with *design thinking* (Cross 2006; Brown 2008) and a plethora of prominent design philosophers use pragmatism to discuss designerly thinking (Buchanan 1992; Dorst and Cross 2001). Thus, even the papers which do not explicitly discuss Dewey, in many ways reflect this empiricist attitude to the world. The ecological perspective of Dewey's pragmatism has led Garrison (Garrison 2009) to describe Dewey's pragmatism as one of constant trans-action. Every living creature is a process, and constantly involved in processes that affects each other. This leads to a strong focus on emergent properties of situations – since every change affected by different processes affects the others, situations will constantly develop and get transformed by the actors in the situation, who strive to achieve *functional co-ordination* with their immediate surroundings. This resonates well with considering design processes as a constant and iterative development of ideas (Dorst and Cross 2001; Wiltschnig, Christensen and Ball 2013) and with a focus on design processes..

## 5.1 Key pragmatic concepts

Throughout my papers I have developed Dewey's philosophy and used it for analyzing design processes. Key concepts have been *problematic situations; inquiring strategies* and transformation, appearing in iteratively developed versions in papers 1, 4 and 7.

### *Problematic situations*

From a Deweyan perspective, we are always engaged in a rhythmic exchange with our environment (Dewey 1934). Existence is considered as a flow, in which we take part, affecting and getting affected all the time. Existence manifests as experiences, most of which most do not stand out and give rise to inquiry. However, at times we find ourselves in problematic situations – something in our environment is not satisfactory, hard to understand, unresolved or intriguing. Thus we are always in transaction with our environment – when a problematic situation is recognized it is the fact that the rhythm of our existence is somehow out of tune with our surrounding environment (Garrison 2009). As we are just one of many such existing processes in a given situation, this out of tune-ness might be due to many different changes, but the main point being that there is now a desire as well as need to recover the balance with our immediate environment. In life in general and in design in particular we might place ourselves in such situations of resistance by choice (Gedenryd 1998). Designers know that design is a process that requires them to make choices in situations which does not have any right or wrong answers – design deals with wicked problems (Buchanan 1992). The idea of wicked problems resonates well with pragmatism, in that pragmatism does away with right or wrong answers as set categories. Rather a solution to a problematic situation can be good or bad, dependent on whether it helps restore the rhythm with the immediate environment.

## *Inquiring strategies*

Inquiring strategies are the approaches taken towards resolving a problematic situation, and they operate on a principle of inference leading to judgment. Inference means to derive logical conclusions from premises that are known or, crucially, assumed to be true, and to Dewey, inference is how we arrive at judgments by assuming something to be true for the purposes of inductive reflection. We recognize a problematic situation and generate potential solutions (induction) and test, mentally or in the world, their implications (deduction). To Dewey, inference is the basic human mode of inquiry: thinking is a "double movement of reflection", a constant back and forth between induction and deduction (Dewey 1910). Dewey distinguishes facts and meaning in a process of inquiry – facts are what is observable and set, whereas meaning is the connections between facts that evolves once we start to resolve a problematic. This process of back and forth is the transformation, and how we progress through a problem. To the trained thinker this process is systematic; it is the gradual organization of facts and meanings, from isolated and fragmentary to a coherent whole (Dewey 1910), meaning that inference and judgment is a skill that can be learned and studied.

When problematic situations occur we enter into a mode of inquiry in which we, through exchange with our environment, try to resolve the problematic situation through a gradual and experimental transformation of the situation's constituent elements (Dewey 1938). By doing so we progress "through" a situation by applying inquiring strategies. In a pragmatist view this approach consists both of what you do as well as the rationale for it. We change our environment in order to satisfy and resolve whatever problematic situation we have encountered by re-establishing the functional coordination (Garrison 2009) with our surroundings. In design this resolving of unsatisfying situations encompass both design thinking and action. Design thinking (Cross 2006) or design judgment (Lera 1981; Nelson and Stolterman 2003) means, in a pragmatist perspective, reformulating one's appreciation of a problematic situation, seeing and appreciation the situation from a new perspective, as well as making a decision on how to proceed. Inquiry in a design thinking perspective is thus the appreciative aspect of transforming and understanding a problematic situation, as well as formulating hypothesises for its resolution. However inquiry might as well be directed at external conditions – at functional coordination with ones physical surroundings. This is crucial to understanding the interplay of design materials and design thinking: In a pragmatist view of resolving problematic situations through inquiring strategies, any distinction between internal (intellectual) or external (physical) transformations is a purely methodological one, rather than a metaphysical one (Garrison 2009). This means that participants in a design progress through a problematic situation by developing and transforming both their understanding of the situation as well as the different materials available in the situation. This is what Schön hints at with his concept of *seeing-moving-seeing* (Schön 1983; Bamberger and Schön 1983; Schön 1992) where each act of seeing and moving refers to transformations of, respectively, ones own perception of a problem (seeing), the external qualities of the situation (moving), and then ones re-evaluation of the problem (seeing).

## *Transformation*

Inquiry unfolds through formations and tests of hypotheses – in this way problematic situations are transformed. In a Deweyan perspective, problematic situations and inquiring strategies are constituted and resolved by transformations supported by technology. Although technology is a broad concept in pragmatism – it encompasses anything used for transforming a situation including theoretical constructs (Hickman 1992) – in my work it has been used to discuss a particular kind of technology, materials in participatory design processes. In pragmatism, technology, and thereby materials constitute the problematic situation but are also at the same time part of the inquiry aimed at resolving the problematic situation. To be precise: materials forms part of a problematic situation, and might also serve as the means of transforming the situation. In action, materials give shape and direction to the problematic situation, but are also shaped and used in mediating inquiry. Gedenryd (Gedenryd 1998) highlights how this can be seen as making the world part of the cognition, drawing our attention to how thinking and doing are intertwined. In such a perspective, viewing materials as pragmatic technology allows us to analyze how materials help designers resolve problematic situations through inquiring strategies – we focus on which role specific materials (technology) played in different design phases (problematic situations).

## 5.2   Design materials as pragmatic technology

Objects, theories and materials used to aid inquiry may be considered 'technology in a pragmatist sense of the word', and regarding materials as technology means recognising how materials constitute both part of the problem and of the solution. "Humans are as much *homo faber*, the tool-making human, as [they are] homo sapiens" (Hickman 1992), referring here to their ability to utilise and adopt technology in an inquiry. This prompts the question: how do different materials support different kinds of transformations as part of design practice? How do materials support inquiry in design?

Consider the prototypical sketching session with the designer seated at a table before receiving a design brief, which he is to resolve, aided by pen and paper. The design brief may be to design a bin for public transportation, like the example used by Dorst and Cross (Dorst and Cross 2001), or Schön's architectural practice example (Schön 1983), mentioned in the introduction of this dissertation. From a pragmatist perspective, the *problematic situation* encompasses everything currently part of the experience of the designer. That means that in addition to the design brief, the pen and paper are also part of the problematic situation, as are the experience and knowledge of the designer, the design brief and so on. The materials form part of the problematic situation simply because they are there – if they are picked up, they may be utilised as part of an *inquiring strategy*, beginning with a sketch. Each line drawn and each reflection is a gradual *transformation* of the problematic situation – experimental, because each line drawn and each reflection are first inductive ('might this work?') and then deductive ('doing this would have these consequences'). This process of inquiry is by no means linear. One induction-deduction loop of drawing a part of a product might lead the designer to realise that his current line of thought is problematic – the next transformation is not of the sketch, but of the designer's own understanding of the problem. This constant back and forth between materials and designer is what Schön designates *seeing-move-seeing* (Schön 1983; Bamberger and Schön 1983; Schön 1987, 1992). But when examining materials in participatory design processes, the emphasis is not on *reflection-in-action* as Schön would have it, but on *transformation* as part of *inquiry*. What becomes important is which transformation it is possible for the designer, or group of designers, to affect at a given time. From a pragmatist perspective, transformations of the internal (the mind) and the external environment are considered symmetrically. And what Dewey refers to as 'the march of the surrounding things' goes on, as each transformation of the world feeds into another.

Pragmatism requires skill to use technology to test inferences, and in design we often use materials for this purpose, for instance, creating prototypes to deduce whether a certain inductive idea of a product is useful and feasible. And we draw on sources of inspiration to present ideas for potential inductive moves. Furthermore, these materials may be transformed – we do not inductively 'have an idea' which is then prototyped, and then, through deduction, rejected or accepted. Some prototypes may be changed on the fly, for instance by writing, or using code to generate several different iterations during a design activity. In the same way, when working with a material such as video, we might utilise the transformability of the material itself. The totality of the situation may be changed, and the design materials offer both the materialisation of a design question as well as a solution.

Based on the preceding discussion of materials as technology for testing inferences, I propose several statements regarding design materials as pragmatic technology.

*Inquiry advances over time, and materials aid this process.* Drawing on Dewey's work, I frame materials as pragmatic technology in several papers (Hansen and Halskov 2014; Halskov and Hansen 2015; Hansen and Halskov 2016). By doing so, I am able to ask how this specific kind of technology helps inquiry to advance, and understand design thinking as systematic inference. Furthermore, as discussed in greater detail in paper 7 (Hansen and Halskov 2016), we may use pragmatism to question how materials help to frame design problems and solutions, and build bridges between them. I have also developed the pragmatist framework to understand how experiences and understanding emerge in a dynamic interplay between a specific technology (e.g. tangible tabletops) and the problematic situation of the user (Hansen 2014). Whereas the initial problematic situation is one of doubt about the nature of the interface, later problematic situations utilise the possibilities of the interface for inquiry. This demonstrates how inquiry moves forward – one transformation feeds into another, the 'march of the surrounding things' (Dewey 1934) is dynamic, and inquiry evolves in leaps and bounds.

*Materials are part of the problematic situations* faced throughout a design process, either supporting design thinking, or as the materials from which we craft products. That makes design materials part of the design problem (Doordan 2003), yet, depending on the *situating strategies* (Gedenryd 1998) we use, different materials may be in focus at different times. We use situating strategies to establish or set up specific problematic situations, because we judge that we need to explore a specific part of a design inquiry. For instance, if we are interested in the general idea with less focus on its specific shape, we might set up a sketching session – the marks-on-paper allow us to transform and develop ideas that may be expressed in that medium (according to our skill), so the part of the design idea we can explore and transform is simultaneously afforded and constrained by the qualities of the materials. Much of the early design work had this flow of design processes and materials: we first developed the general idea from a vague form to a more specific one, through brainstorming, sketching and finally into a prototype. In the process of development the materials that were part of the final product came more and more into view through a gradual refinement. Starting with marks on paper means that some aspects of the problematic situation may be transformed, but not others – we may get an overall idea, but we are unable to transform our perception of the material qualities of the design – but at the same time, it is difficult to use transformations of materials as inspiration for transformations of our design idea. Since the materials are not readily available as technology in a pragmatist sense (because they are not part of this design session), such transformations of the design idea are left for later. In a counter-example, starting by exploring materials means conducting *materials-driven* design, reflecting a bricolage approach (Vallgårda and Fernaeus 2015). Since materials are not only part of the design problem (since they have to be incorporated into the final design), but also are part of the immediate problematic situation, beginning with the materials means that the initial design space becomes constrained at a specific point.

*Materials constrain a design space*, however, design constraints are not necessarily a bad thing – rather, they may function in a *creative* role, as detailed by Biskjær and Dalsgaard (Biskjær and Dalsgaard 2012), and indeed, most design methods operate on the principle of constraining a design space. Constraints are a part of the situating strategy that leads to successful inquiry, because it gives direction and shape to the transformations that unfolds as inquiry progresses. By drawing in and selecting materials, we simultaneously constrain a design space (since participants in a design activity will use the materials) and offer potential transformations.

*Materials give direction and shape to an inquiry.* The selection of materials in a situating strategy means that we have selected a direction of inquiry and an area of the design space to be explored, rather than a solution. Since no inquiry starts in a vacuum, what we do first matters – each problematic situation is part of the 'march of existence', as Dewey put it, meaning that if we look at design processes as a gradually-evolving string of problematic situations, we can choose the focus of the design process through our choice of materials in the problematic situations we set up for ourselves, and what we choose to draw into the ones into which we are thrown, among many other things.

# 6. Discussion

In the previous chapters I outlined my work and positioned it in the field of interaction design. In this chapter I summarise and discuss my research contributions from the papers appended at the end of this overview article, in relation to the topics of *design materials*, *design processes* and *participatory design*.

Pragmatism forms the theoretical framework for understanding how materials help the development of ideas in design processes in participatory and interaction design. I have highlighted how my work may be seen as a way of answering Rogers' call encouraging HCI researchers to identify and develop a theoretical concept and framework that 'provides conceptual tools and a cogent set of arguments or propositions that can explain or articulate phenomena' (Rogers 2012). My contributions are what Rogers defines as *descriptive* and *explanatory*. Descriptive, because they provide concepts for describing and analysing design practice, enabling further questions to be raised; explanatory, because my contributions explicate the relationships and processes going on when using materials in design work.

## 6.1 Design materials

My work has focused on design materials, and proposes that we consider them part of the problematic situation faced in design work. By doing so, I have argued for a symmetrical understanding of the materials used for design thinking and for shaping products. I have done so because when we use materials in design thinking, we use materials to explore problems and solutions. In my paper on materials in creative design processes (Hansen and Halskov 2016), I argue that we might see the design process as a series of *problematic situations* constituted by, and resolved through, the use of materials. Some of these materials are for design thinking, whereas others, through the process, become part of the final product.

My work highlights how materials might form the point of departure for design ideas, as discussed by Wiberg (Wiberg 2013) and Vallgårda and Fernaeus (Vallgårda and Fernaeus 2015). Starting from the specific material properties of the possibilities of the pixels of a media facade under development (Hansen and Halskov 2016), my work shows how design materials have a broader role than previously highlighted in *materials selection* (Ashby and Johnson 2013). In the example from (Hansen and Halskov 2016), the materials in the mock-up are used for both design reflection, setting the stage for further ideation, and to carry out concrete material experiments. As a contrast of materials used for pure design reflection without any reference to the final product, my early work on using physical materials in participatory design processes (Hansen and Dalsgaard 2012) shows the value of framing participant understanding of the matter at hand through a collaborative process. The physical materials – pen, paper, cardboard and Post-its® – are used to document and transform the design space.

My work explores how materials shape the use of a product. A key point is that the material qualities of a product – here, in the form of a tangible tabletop – establish the point of departure for a material exploration of form and sound (Hansen and Halskov 2014). By drawing on pragmatism, my work may be seen engaging with the work of Vallgårda et al. (Vallgårda and Fernaeus 2015), showing how pragmatism might be used to understand engagement with computational composites that utilise the possibilities for changes in temporal form.

New materials give rise to new interfaces and design possibilities. The two media facade cases included in my work (Korsgaard et al. 2012; Hansen and Halskov 2016) are examples of this, since many of the challenges they present are novel. The diversity of available materials means there is a need for both exploration of what materials and materialities we might imagine (Robles and Wiberg 2010, 2011), and for design process perspectives on designing with these materials, which I have shown with the cases presented. In a way, the contrast between the many brilliant reflections on what materials and materiality mean and my design process perspective, is emblematic of this development. Both perspectives contribute, and both concern the questions of what design is and can be.

Finally, crafting was a focus in a paper where I specifically explored Sennett's work (Sennett 2008), to discuss how coding might be understood as a craftsman's engagement with the code running on an old AMIGA computer (Hansen, Nørgård and Halskov 2014). In doing so, I showed how code might be considered a material that may be manipulated and drawn on throughout a design process. Although the focus on crafting is not explicit in all my papers, the perspective of crafting and its implicit focus on a conversation with materials resonates with pragmatism. Sennett's work (Sennett 2008), on which much of the craftsmanship strand of interaction design and HCI literature draws, resonates with Dewey's concept of systematic inference (Dewey 1910), and I advance as a useful way of understanding design reflection. From a pragmatist view of craftsmanship, each change to the code is the result of an induction that is tried out deductively in the real world. Through judgment (Nelson and Stolterman 2003), we decide whether the transformation of the code is desirable, thus revealing the constant exchange between an individual and its environment.

## 6.2 Design processes

My work focuses on developing interaction design research by contributing theoretical concepts for understanding materials. In chapter 3, I defined a set of salient aspects of design processes that my work has focused on: how design processes concern experimentally working through wicked problems; how methods add to a designer's repertoire by crystallising the insights of design research into a form that designers can adopt; and how such applications depend on the ability of the designer to reflect on the unique situation. By crystallising insights of my own design research, I have contributed a set of examples of media architecture design processes (Korsgaard et al. 2012) on which others may draw, and integrate into their design repertoire.

Materials support the development of ideas. Although the key to design thinking is the ability to reflect, judge and select the best options for resolving a dilemma, design processes are not just a mental exercise. A host of materials supports the development of ideas. In chapter 4, I discussed this specifically with regard to sketching, prototyping and external representations, where I highlighted how these materials helped to frame problems and try out potential solutions to them. As one example, I developed a conceptual framework (Hansen and Dalsgaard 2012) for understanding how physical materials support design work by aiding collaborative reflection. By drawing on pragmatism, our attention is drawn to how a shared design space is manifested and transformed through collaborative acts of transforming materials. Crystallising the insights from this analysis as design considerations is one way to share such insights with the wider design research community. Apart from highlighting and demonstrating the value of pragmatism, the specific design considerations (rapid transformations; documenting decisions; aligning collaborative efforts; provoking reflection; proposing and supporting design changes) are also potential 'reflections-in-action' that others who use physical materials in workshops may draw on directly, without necessarily delving as deeply into Dewey's ideas as this dissertation has done.

Another example of how materials aid the development of design ideas is provided in (Hansen and Halskov 2016). Here, a much wider array of materials is examined, in that we use materials both for design thinking and for shaping the product in a constant back and forth between these two roles. By casting the problematic situations of the EXPO 2010 case as concerned with the specific materials at play at any given point in the design process, I have tried to show how Doordan's (Doordan 2003) point about the 'materials being part of the design problem' may have wider implications. Designer's do not choose a function and select a material based on it, as function-driven materials selection would have it. Rather, as the experiments with the Mock-up, Pixel-Tool and Mixed reality model showed (Hansen and Halskov 2016), design problems are always, throughout the entire design process, deeply intertwined with their current material manifestations. Dorst and Cross drew our attention to how, in design processes, problems and solutions co-develop (Dorst and Cross 2001). However, this co-development is supported by materials, and taking pragmatism seriously dissolves the distinction between 'the idea' and its 'manifestation' in the material. There is just the problematic situation, which consists of both the people participating in the design process and the materials they draw on to resolve the situation.

Last I have explored a crafting perspective on design processes. Rather than the function-driven approach of much interaction design (Wiberg 2013, 2015), in (Hansen, Nørgård and Halskov 2014) I have highlighted how we might see the process of coding digital art as a form of crafting. By doing so, I have attempted to offer a perspective that cuts across dichotomies of art and engineering.

## 6.3 Participatory Design

Participatory Design constitutes a specific perspective on design that has formed the background and context for my work. I previously defined PD as concerned with establishing situations of democratic dialogue and mutual learning through design events. By doing so, PD aims to enable participants and potential users to have a voice in the design process. My main contributions to PD concern identifying what constitutes PD, and tackling two separate issues of PD research: materials may be used to involve participants in design; and 'participation' is a complex phenomenon that must be negotiated and analysed on a case-by-case basis.

How we might define PD is a big question, but it is a question that we must constantly ask ourselves as the field develops and progresses. In a literature review (Halskov and Hansen 2015), my co-author and I identified the existence of a great diversity of definitions and practices of Participatory Design. I also was able to identify five archetypical types of PD contributions, and reformulate the key aspects of contemporary PD as politics, people, methods, context and products. Such work has value, precisely because it indicates a direction, both consolidating the state of art of the field, and identifying issues that need attention. PD is emblematic of the conditions under which all design practice operates. It interplays with the surrounding society, with its changing technologies and organisational contexts. As such, PD methods developed in the infancy of the field must be constantly renegotiated, leading me to suggest that none of the aspects of PD may be considered on its own. Of necessity, all PD research concerns the combination of two or more of those aspects. For instance, it makes little sense to ask what a 'PD method' is – what matters is how that method is used in a specific situation, for instance, to raise political issues, engage with new domains or to prototype technology.

One point that emerged from the literature review (Halskov and Hansen 2015) was that what constituted 'participation' is a rather complex question. Although one might be tempted to suggest that 'participation' means 'being part of the workshop', such a formulation, although not incorrect, also hides the fact that a lot of design processes takes place outside of workshops. And the content of the workshop was negotiated before the workshop participants arrived, in a planning phase. This doubt about the binary nature of participation gave rise to an analysis based on a case study of how participation might appear from an Actor-Network Theory perspective. In this journal article (Andersen et al. 2015) we used the theoretical concept of a 'matter of concern' as opposed to a 'matter of fact'. A matter of concern is something that cannot be exhaustively defined before a concrete situation appears – we must always make choices in the specific situation, and then account for the bounded rationale of those choices, to the best of our ability. Therefore, in this article we argue that we must dissolve the binary idea of participation: participants are networks, rather than individual subjects. Accounting for our choices in a specific design situation is what constitutes proper PD practice.

# 7. Conclusion

I started this overview article from the question of how we might understand materials in interaction design processes. This question have been researched over three years using a research-through-design approach in which I have engaged directly with design processes in order to unravel this question. Through describing and discussing my work as taking place within a specific *experimental* system (Rheinberger 1997, 2010; Dalsgaard 2016) I have tried to describe the context in which my work have unfolded. This experimental system has offered me a frame with regards to questions and interests where much of my work has interlocked with that of my colleagues. Furthermore the experimental system offered me the facilities and technical expertise to conduct my work in. At the same time my work has been part of the experimental system, and taking part in shaping the work in the research projects I have joined.

Within the fields of interaction design and HCI, there are several strands of interest in materials: computers as material things, materials for design thinking and crafting. None of these three perspectives reflects attempts to ontologically fix what we might mean when we talk about 'materials'. Instead, they reflect different theoretical points of departure, and more importantly, they reflect different yet complimentary research interests. Of necessity, engaging with practical design work means engaging in ways that open one's eyes to the value of all three perspectives, at different times, and for different purposes. In one of my papers (Hansen and Halskov 2016), I straddle this gap by showing how materials and ideation are deeply intertwined, but how we might also choose, as an exercise in design skills, to put down the materials of the product for a while, and sketch a bit in another material. Such a view of design reflects the empiricist attitude to the world that is also part of pragmatism. Although frustrating to developing designers, that also reflects how skill and experience with materials ultimately colour the choices made in a design process. Through the work of Schön, and through design thinking, pragmatism already has a strong and enduring influence on how design practice is discussed academically, and by drawing on the pragmatist concepts advanced here and in the appended papers, different perspectives on design materials may be brought together under the umbrella of pragmatism.

# 9. Appended Papers

Faculty of Arts, Aarhus University

## Statement of co-authorship

A statement of co-authorship is required for each academic publication which has been produced in collaboration with other people and which is to be assessed in connection with your application. Please upload such statements of co-authorship as appendices to your application.

Co-authors are defined in accordance with the Vancouver requirements, cf. http://www.icmje.org/index.html

This statement of co-authorship applies to the following article/work:

**The Productive Role of Material Design Artefacts in Participatory Design Events**

The amount which **Nicolai Brodersen Hansen** has contributed to the article/work is calculated based on the following scale:

A. The applicant has made some contribution (0-33%)
B. The applicant has made a significant contribution (34-66%)
C. The applicant has (to a large extent) done the work independently (67-100%)

| Statement regarding individual elements | A,B,C |
| --- | --- |
| 1. Formulating the basic academic issue (at the ideas stage) based on theoretical questions for which answers are required, including the synthesis of the issue into questions deemed to be answerable by carrying out analyses or concrete experiments or studies. | C |
| 2. Planning experiments/analyses and drawing up a study methodology in such a way that it is reasonable to expect that the questions posed under point 1 can be answered, including the choice of methods and independent method development. | C |
| 3. Being involved in analysis work or concrete experiments/studies. | C |
| 4. Presenting, interpreting and discussing the results achieved in the article. | B |

Signatures of co-authors

| Date | Name | Title | Signature |
| --- | --- | --- | --- |
| 04-05-2016 | Peter Dalsgaard | Associate Professor | 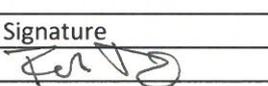 |
| | | | |
| | | | |
| | | | |

Signature of applicant:

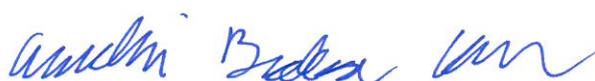

# The Productive Role of Material Design Artefacts in Participatory Design Events

Nicolai Brodersen Hansen and Peter Dalsgaard
CAVI & PIT Aarhus University imvnbh@hum.au.dk, dalsgaard@cavi.dk

## Abstract
Physical design artefacts are employed in a wide range of participatory design events, yet there are few comprehensive discussions of the properties and qualities of them in the literature of the field. In this paper, we examine *the productive role of material design artefacts in participatory design events*. First, we offer a theoretical foundation for understanding material artefacts in design, based on pragmatist philosophy. Then, we employ this theoretical perspective to analyse a case in which a range of physical design materials was employed to envision and explore a projected building, the "Urban Media Space" a new library in Aarhus, Denmark. Drawing on examples from this case, we define a series of design considerations for employing material design artefacts in collaborative design events. Our contribution is valuable both in advancing the theoretical standpoint of interaction design in general, and in allowing participatory design practitioners to reflect on their use of material design artefacts when involving users.

## Citation format
Nicolai Brodersen Hansen and Peter Dalsgaard. 2012. The productive role of material design artefacts in participatory design events. In *Proceedings of the 7th Nordic Conference on Human-Computer Interaction: Making Sense Through Design* (NordiCHI '12). ACM, New York, NY, USA, 665-674. DOI=http://dx.doi.org/10.1145/2399016.2399117

## Link
http://dl.acm.org/citation.cfm?id=2399117

Faculty of Arts, Aarhus University

## Statement of co-authorship

A statement of co-authorship is required for each academic publication which has been produced in collaboration with other people and which is to be assessed in connection with your application. Please upload such statements of co-authorship as appendices to your application.

Co-authors are defined in accordance with the Vancouver requirements, cf. http://www.icmje.org/index.html

This statement of co-authorship applies to the following article/work:

**Odenplan: a media façade design process**

The amount which **Nicolai Brodersen Hansen** has contributed to the article/work is calculated based on the following scale:

A. The applicant has made some contribution (0-33%)
B. The applicant has made a significant contribution (34-66%)
C. The applicant has (to a large extent) done the work independently (67-100%)

| Statement regarding individual elements | A,B,C |
|---|---|
| 1. Formulating the basic academic issue (at the ideas stage) based on theoretical questions for which answers are required, including the synthesis of the issue into questions deemed to be answerable by carrying out analyses or concrete experiments or studies. | B |
| 2. Planning experiments/analyses and drawing up a study methodology in such a way that it is reasonable to expect that the questions posed under point 1 can be answered, including the choice of methods and independent method development. | B |
| 3. Being involved in analysis work or concrete experiments/studies. | B |
| 4. Presenting, interpreting and discussing the results achieved in the article. | B |

Signatures of co-authors

| Date | Name | Title | Signature |
|---|---|---|---|
| 04-05-2016 | Henrik Korsgaard | Phd Student | 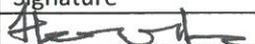 |
| 04-05-2016 | Ditte Amund Basballe | Postdoc | 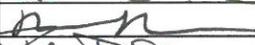 |
| 04-05-2016 | Peter Dalsgaard | Associate Professor | 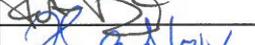 |
| 04-05-2016 | Kim Halskov | Professor | 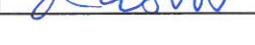 |

Signature of applicant:

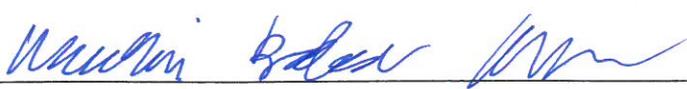

# Odenplan: a media façade design process

Henrik Korsgaard, Nicolai Brodersen Hansen, Ditte Basballe, Peter Dalsgaard and Kim Halskov

## Abstract
In this paper we present an example of how to work with the challenges inherent in media façade design processes. We base the paper on our experiences from the creation of a series of design proposals for a media façade on the Odenplan subway station in Stockholm, Sweden. We approach the question of how to design for media façades by discussing how we have structured our design process to address specific sets of challenges outlined in previous literature in the field of media architecture. In our view, such research is valuable in that it helps establish common ground for researchers and practitioners in a developing field by building a repertoire of approaches, as well as highlight important issues that need to be addressed in this emergent field.

## Citation format
Henrik Korsgaard, Nicolai Brodersen Hansen, Ditte Basballe, Peter Dalsgaard, and Kim Halskov. 2012. Odenplan: a media façade design process. In *Proceedings of the 4th Media Architecture Biennale Conference: Participation* (MAB '12). ACM, New York, NY, USA, 23-32. DOI=http://dx.doi.org/10.1145/2421076.2421081

## Link
http://dl.acm.org/citation.cfm?id=2421081

Faculty of Arts, Aarhus University

## Statement of co-authorship

A statement of co-authorship is required for each academic publication which has been produced in collaboration with other people and which is to be assessed in connection with your application. Please upload such statements of co-authorship as appendices to your application.

Co-authors are defined in accordance with the Vancouver requirements, cf. http://www.icmje.org/index.html

This statement of co-authorship applies to the following article/work:

**Crafting code at the demo-scene**

The amount which **Nicolai Brodersen Hansen** has contributed to the article/work is calculated based on the following scale:

- A. The applicant has made some contribution (0-33%)
- B. The applicant has made a significant contribution (34-66%)
- C. The applicant has (to a large extent) done the work independently (67-100%)

| Statement regarding individual elements | A,B,C |
|---|---|
| 1. Formulating the basic academic issue (at the ideas stage) based on theoretical questions for which answers are required, including the synthesis of the issue into questions deemed to be answerable by carrying out analyses or concrete experiments or studies. | B |
| 2. Planning experiments/analyses and drawing up a study methodology in such a way that it is reasonable to expect that the questions posed under point 1 can be answered, including the choice of methods and independent method development. | C |
| 3. Being involved in analysis work or concrete experiments/studies. | B |
| 4. Presenting, interpreting and discussing the results achieved in the article. | B |

Signatures of co-authors

| Date | Name | Title | Signature |
|---|---|---|---|
| 04-05-2016 | Rikke Toft Nørgård | Assistant Professor | |
| 04-05-2016 | Kim Halskov | Professor | |
| | | | |
| | | | |

Signature of applicant:

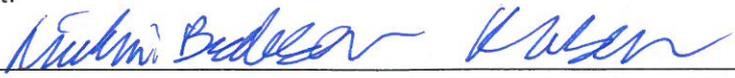

# Crafting Code at the Demo-scene

Nicolai Brodersen Hansen, Rikke Toft Nørgård and Kim Halskov
CAVI, PIT and TDM. Aarhus University.
nbhansen@cavi.au.dk, rtoft@tdm.au.dk, halskov@cavi.au.dk

Abstract

This paper introduces the idea of craftsmanship as a way of understanding the shaping and re-shaping of code as a material crafting practice. We build our analysis on a qualitative study of a coder engaged in creative and expressive programming on an old hardware platform. The contribution of the paper is a set of conceptual categories: *craft engagement*, *craftsmanship rhythm* and *craftsmanship expressivity*, that conceptualizes coding as crafting.

Faculty of Arts, Aarhus University

## Statement of co-authorship

A statement of co-authorship is required for each academic publication which has been produced in collaboration with other people and which is to be assessed in connection with your application. Please upload such statements of co-authorship as appendices to your application.

Co-authors are defined in accordance with the Vancouver requirements, cf. http://www.icmje.org/index.html

This statement of co-authorship applies to the following article/work:

**Material Interactions with Tangible Tabletops: a Pragmatist Perspective**

The amount which **Nicolai Brodersen Hansen** has contributed to the article/work is calculated based on the following scale:

A. The applicant has made some contribution (0-33%)
B. The applicant has made a significant contribution (34-66%)
C. The applicant has (to a large extent) done the work independently (67-100%)

| Statement regarding individual elements | A,B,C |
|---|---|
| 1. Formulating the basic academic issue (at the ideas stage) based on theoretical questions for which answers are required, including the synthesis of the issue into questions deemed to be answerable by carrying out analyses or concrete experiments or studies. | B |
| 2. Planning experiments/analyses and drawing up a study methodology in such a way that it is reasonable to expect that the questions posed under point 1 can be answered, including the choice of methods and independent method development. | B |
| 3. Being involved in analysis work or concrete experiments/studies. | B |
| 4. Presenting, interpreting and discussing the results achieved in the article. | B |

Signatures of co-authors

| Date | Name | Title | Signature |
|---|---|---|---|
| 04-05-2016 | Kim Halskov | Professor | 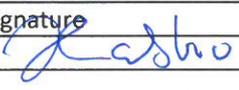 |
| | | | |
| | | | |
| | | | |

Signature of applicant: 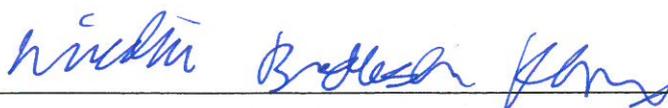

# Material interactions with tangible tabletops: a pragmatist perspective

Nicolai Brodersen Hansen and Kim Halskov

## Abstract
We investigate how the interaction with tangible interactive tabletops can be seen as a material exploration of form and sound. As the theoretical foundation for our analysis we build on John Dewey's pragmatism as well as recent efforts to appropriate pragmatism for interaction design research. As the research platform for this investigation we developed an interactive tabletop, the Radar Table, which allows users to create soundscapes by manipulating tangible objects. The Radar Table was deployed 'in the wild' at a major Danish music festival, and based on video recordings we examine people's dynamic exploration of sound through the interactive tabletop. The main contribution of the paper is the development of the theoretical foundation for understanding tangible tabletops as material interfaces that can be shaped and experimented with. We build on three of the basic concepts of pragmatism: situation, inquiry, and technology, which we develop further for the study of the dynamics of material interactions with tangible tabletops as part of a research strategy of appropriating pragmatism for use in interaction design and HCI research.

Faculty of Arts, Aarhus University

## Statement of co-authorship

A statement of co-authorship is required for each academic publication which has been produced in collaboration with other people and which is to be assessed in connection with your application. Please upload such statements of co-authorship as appendices to your application.

Co-authors are defined in accordance with the Vancouver requirements, cf. http://www.icmje.org/index.html

This statement of co-authorship applies to the following article/work:

**The diversity of participatory design research practice at PDC 2002–2012.**

The amount which **Nicolai Brodersen Hansen** has contributed to the article/work is calculated based on the following scale:

- A. The applicant has made some contribution (0-33%)
- B. The applicant has made a significant contribution (34-66%)
- C. The applicant has (to a large extent) done the work independently (67-100%)

| Statement regarding individual elements | A,B,C |
|---|---|
| 1. Formulating the basic academic issue (at the ideas stage) based on theoretical questions for which answers are required, including the synthesis of the issue into questions deemed to be answerable by carrying out analyses or concrete experiments or studies. | B |
| 2. Planning experiments/analyses and drawing up a study methodology in such a way that it is reasonable to expect that the questions posed under point 1 can be answered, including the choice of methods and independent method development. | B |
| 3. Being involved in analysis work or concrete experiments/studies. | B |
| 4. Presenting, interpreting and discussing the results achieved in the article. | B |

Signatures of co-authors

| Date | Name | Title | Signature |
|---|---|---|---|
| 04-05-2016 | Kim Halskov | Professor | 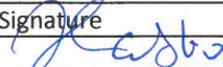 |
| | | | |
| | | | |
| | | | |

Signature of applicant: 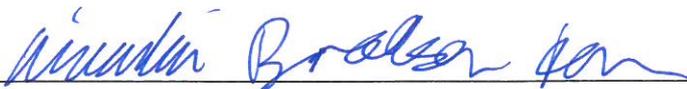

# The diversity of participatory design research practice at PDC 2002–2012

Kim Halskov and Nicolai Brodersen Hansen
halskov@cavi.au.dk, nbhansen@cavi.au.dk

## Abstract
We investigate the diversity of participatory design research practice, based on a review of ten years of participatory design research published as full research papers at the Participatory Design Conferences (PDC) 2002–2012, and relate this body of research to five fundamental aspects of PD from classic participatory design literature. We identify five main categories of research contributions: Participatory Design in new domains, Participatory Design methods, Participatory Design and new technology, Theoretical contributions to Participatory Design, and Basic concepts in Participatory Design. Moreover, we identify how participation is defined, and how participation is conducted in experimental design cases, with a particular focus on interpretation, planning, and decision-making in the design process.

## Citation format
Halskov, Kim, and Nicolai Brodersen Hansen. "The diversity of participatory design research practice at PDC 2002–2012." *International Journal of Human-Computer Studies* 74 (2015): 81-92.

## Link
http://www.sciencedirect.com/science/article/pii/S1071581914001220

Faculty of Arts, Aarhus University

## Statement of co-authorship

A statement of co-authorship is required for each academic publication which has been produced in collaboration with other people and which is to be assessed in connection with your application. Please upload such statements of co-authorship as appendices to your application.

Co-authors are defined in accordance with the Vancouver requirements, cf. http://www.icmje.org/index.html

This statement of co-authorship applies to the following article/work:

**Participation as a matter of concern in participatory design**

The amount which **Nicolai Brodersen Hansen** has contributed to the article/work is calculated based on the following scale:

- A. The applicant has made some contribution (0-33%)
- B. The applicant has made a significant contribution (34-66%)
- C. The applicant has (to a large extent) done the work independently (67-100%)

| Statement regarding individual elements | A,B,C |
|---|---|
| 1. Formulating the basic academic issue (at the ideas stage) based on theoretical questions for which answers are required, including the synthesis of the issue into questions deemed to be answerable by carrying out analyses or concrete experiments or studies. | B |
| 2. Planning experiments/analyses and drawing up a study methodology in such a way that it is reasonable to expect that the questions posed under point 1 can be answered, including the choice of methods and independent method development. | A |
| 3. Being involved in analysis work or concrete experiments/studies. | A |
| 4. Presenting, interpreting and discussing the results achieved in the article. | B |

Signatures of co-authors

| Date | Name | Title | Signature |
|---|---|---|---|
| 04-05-2016 | Lars Bo Andersen | Postdoc | *signed* |
| 04-05-2016 | Peter Danholt | Associate Professor | *signed* |
| 04-05-2016 | Peter Lauritsen | Professor | *signed* |
| 04-05-2016 | Kim Halskov | Professor | *signed* |

Signature of applicant: *signed*

# Participation as a matter of concern in participatory design

Lars Bo Andersen, Peter Danholt, Kim Halskov, Nicolai Brodersen Hansen and Peter Lauritsen.

## Abstract

This article starts from the paradox that, although participation is a defining trait of participatory design (PD), there are few explicit discussions in the PD literature of what constitutes participation. Thus, from a point of departure in Actor-Network Theory (ANT), this article develops an analytical understanding of participation. It is argued that participation is a *matter of concern*, something inherently unsettled, to be investigated and explicated in every design project. Specifically, it is argued that (1) participation is an act *overtaken* by numerous others, rather than carried out by individuals and (2) that participation *partially exists* in all elements of a project. These traits are explicated in a design project called 'Teledialogue', where the participants are unfolded as networks of reports, government institutions, boyfriends, social workers and so on. The argument is synthesised as three challenges for PD: (1) participants are network configurations, (2) participation is an aspect of all project activities and (3) there is no gold standard for participation.

Faculty of Arts, Aarhus University

## Statement of co-authorship

A statement of co-authorship is required for each academic publication which has been produced in collaboration with other people and which is to be assessed in connection with your application. Please upload such statements of co-authorship as appendices to your application.

Co-authors are defined in accordance with the Vancouver requirements, cf. http://www.icmje.org/index.html

This statement of co-authorship applies to the following article/work:

**(The Role of) Materials in Design Processes**

The amount which **Nicolai Brodersen Hansen** has contributed to the article/work is calculated based on the following scale:

- A. The applicant has made some contribution (0-33%)
- B. The applicant has made a significant contribution (34-66%)
- C. The applicant has (to a large extent) done the work independently (67-100%)

| Statement regarding individual elements | A,B,C |
|---|---|
| 1. Formulating the basic academic issue (at the ideas stage) based on theoretical questions for which answers are required, including the synthesis of the issue into questions deemed to be answerable by carrying out analyses or concrete experiments or studies. | C |
| 2. Planning experiments/analyses and drawing up a study methodology in such a way that it is reasonable to expect that the questions posed under point 1 can be answered, including the choice of methods and independent method development. | C |
| 3. Being involved in analysis work or concrete experiments/studies. | C |
| 4. Presenting, interpreting and discussing the results achieved in the article. | B |

Signatures of co-authors

| Date | Name | Title | Signature |
|---|---|---|---|
| 04-05-2016 | Kim Halskov | Professor | 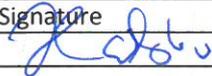 |
| | | | |
| | | | |
| | | | |

Signature of applicant: 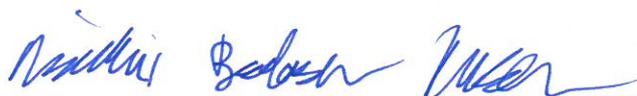

# (The Role of) Materials in Design Processes

Nicolai Brodersen Hansen, Kim Halskov

nbhansen@cavi.au.dk, halskov@cavi.au.dk

## ABSTRACT

Materials are a critical component in design processes yet there are few comprehensive discussions addressing the role of materials in the development of ideas during the design process. We build on Dewey's pragmatist philosophy by framing materials as *pragmatic technology* and introducing and developing the concepts of *situation*, *inquiry,* and *transformation* into a coherent framework usable for understanding the role of materials in creative design processes. As the principal case we investigate the design process of a media façade for the 2010 World Expo in Shanghai. Through this analysis we identify three important roles for materials in design: How does design materials 1) establish *problematic situations*, 2) help resolve them through *inquiring strategies,* and 3) thus support *transformation* of the problematic situation. This pragmatist framework constitutes the main contribution and enables researchers to reflect on what role materials play in creative design processes.

## INTRODUCTION

This paper examines one aspect of design processes - how design ideas evolve through materials in design processes encompassing both designerly thinking (Dorst and Cross 2001) and acting (Gedenryd 1998a). Drawing on Dewey's pragmatist philosophy the paper contributes to the current discussion on materiality and materials in design by analyzing in-depth how materials play different roles in creative design processes.

By developing concepts drawn from Dewey's pragmatism our work build on a well-established school of thought, which previously has demonstrated its value as a foundation for design research, (Schön 1992; Dalsgaard 2014) . Pragmatism allows us to analyze and reflect on the continuous development and refinement of design ideas and materials during the design process. By considering materials as *pragmatic technology* (Hickman 1992) researchers might reflect on how materials matter both as framings for problematic situations and as components in resolving these problematic situations. The pragmatist framework enables researchers to reflect on how ideas and materials co-develop in design processes, which we illustrate through the analysis of the design of the media architecture of the Danish pavilion at the Expo 2010 in Shanghai.

We first outline the differing understandings of materials in IxD and HCI, before moving on to discussing materials in a pragmatist perspective and analyzing the Expo 2010 case. The paper concludes with a discussion of how design materials supported 1) establishing *problematic situations*, 2) help resolve them through *inquiring strategies,* and 3) thus support *transformation* of the problematic situation.

# Materials in interaction design and human-computer interaction

Design is a material praxis – and new materials challenge designers by their novelty but also offer opportunities for design that challenge existing paradigms within HCI and IxD research. Thus, materials and materiality have always played a practical role in interaction design and HCI and recently there has been a strong and growing research interest in materials (Wiberg, Kaye and Thomas 2013; Wiberg 2013; Robles and Wiberg 2010). Yet this research interest in materials and materiality is a recent development. This is because the main focus of interaction design has been WIMP-interface or smart phones, meaning that most design processes dealt with systems using similar material shapes and configuration. The diversity in systems has historically dealt with what happened on the screen (or screens in the case of networked systems) rather than the material qualities of the designed object. Software was thus the material that interaction designers worked with while the rest of the materials were in the background and not up for questioning. Thus much research attention has historically been directed towards designing systems on different kinds of screens on stationary computers, laptops, mobile devices, yet still working with software as the material of WIMP.

But in the past 10 years the advent of new smart materials, the Arduino, LEDs integrated into buildings and countless other advances, have given rise to new forms of digital products that challenge the idea of software as the material of IxD and HCI. Pervasive computing, as well as the emerging field of media architecture provides notable examples (Dalsgaard and Halskov 2010), as does more experimental forms and material configurations (Jung and Stolterman 2012; Vallgårda 2014). Such examples demonstrate how new materials open up new technical avenues for interaction designers to explore and expands design horizons beyond WIMP- and mobile interfaces (Wiberg 2013; Robles and Wiberg 2010) .

By considering HCI and IxD as material practices, it becomes clear that designing interactive systems is not something that is apart from other kinds of design. Rather the new forms and shapes that computational technology can take highlight the interplay between materials and designing. Italian philosopher Ezio Manzini, in his book "The Material of Invention" (Manzini and Cau 1989), mirrors this view of material availability influencing research and design areas. Manzini pointed out how the advent of plastics and other new materials implies that material selection have now become a hyper-selection. There are so many new materials available and the materials can have so many new properties, that designers and engineers face a unique challenge (NO CITED PAGES FOR REPEATED CITATION). Vallgårda and Redström (Vallgårda and Redström 2007) have pointed out this development as also taking place within the design of computational objects in the sense that all computational objects are blends of physical and digital materials. Thus the hyper-selection of materials that Manzini (Manzini and Cau 1989) describes is not confined to plastics but also concerns designing interactive systems.

## Design materials

Designing then becomes a question both of having a material knowledge but also of consciously exploring materials. Such a view of materials aligns well with Schön's conceptualization of designing as a conversation with the materials (Schön 1992), as well as with a view of problems and solutions in design as outlined by Dorst and Cross (Dorst and Cross 2001). To Dorst and Cross it is not the case that creative design processes first define a problem and then look for a solution. Rather the problem and the solution co-evolve, something that has later been elaborated by (Wiltschnig, Christensen and Ball 2013) who highlights how the problem and solution space of a design problem are always entangled. Hence, creative design processes is the conscious exploration of problems and solutions through both thinking and doing. Thus design activities involving materials become something more than the mere construction of ideas through material selection fit to the purpose of a system(Ashby and Johnson 2013). Rather designing must be understood as a constant exchange between thinking and doing, between reflection and shaping of materials and thus see creative design processes as an activity spanning an entire ecology of work.

The question then becomes, how does one articulate and reflect on the process of designing with materials? We suggest pragmatism as a way of analyzing and reflecting on design activity as something that always takes place within a specific situation and environment, rife with opportunity, resistance and inspiration.

Materials in design are used in two ways in design research and practice: both as the material out of which designers craft products but also as materials that are part of the design process alone. Designers use and discard them as externalizations (Dix and Gongora 2011) while developing ideas. For example, designers might use pen and paper to sketch out a proposed solution, each sketch building on the previous as he works through the design space. The successive sketches are temporarily used materials that designers discard at some point in the design process. To reflect on the duality of materials used both for crafting products and developing design ideas we suggest the term *design materials*: those physical and computational things that are either used in crafting the final product or used and consumed as part of the design process. An example of the latter could be the use of prototypes, which are used to explore alternatives, much like an architect uses his sketchpad to try out alternatives for the design of a building. In that case, the material of the sketch stands instead of the building materials, because this offers desirable options in the sketching phase of the work. In other phases the architects and artisans involved in a building project experiments with different materials like concrete, wood and glass, thus examining the qualities of the material proper. To advance the understanding of the role of materials in creative design processes this paper focuses on the way materials support the development and testing of design ideas. Design ideas, their form, function and materials, evolve as a continuous interplay between design thought and action. This emergent relationship leads us to suggest pragmatism as a perspective on design materials.

## A pragmatist view of design materials

Pragmatism offers a powerful foundation for interrogating the interplay of design materials and ideation in creative design processes, because pragmatism focus on thinking and doing as situational activities. Thereby pragmatism allows us to reflect on the ultimate particulars (Nelson and Stolterman 2014) that are individual design situations. At the same time, pragmatism has a long history in design theory, most prominently through the work of Donald Schön (Schön 1983). Recently Dalsgaard (Dalsgaard 2014) have argued that pragmatism and *design thinking* share a common point of departure and that design thinking can be well illuminated through concepts drawn from pragmatism.

Below we outline the pragmatist philosophy perspective and develop a theoretical framework for understanding the roles of design materials. We base ourselves primarily on the philosopher John Dewey, whose work has been developed in design (Rylander n.d.; Melles 2008; Dalsgaard 2014) and which originally informed the pragmatist point of departure of Schön (Schön 1983).

Pragmatism is a praxis-based philosophy in which categories and ideas of absolute truths are dispensed in favor of a situated view in which an idea, theory or object are evaluated based on its usefulness in a specific situation. Dewey's pragmatism is strongly influenced by an ecological perspective. In his book "Art as Experience" (Dewey 1934) Dewey describes human existence as something that is always in a rhythmic exchange with its environment. This has led Garrison (Garrison 2009) to describe Dewey's pragmatism as one of constant *trans-action*. Thereby every living creature is a process, and constantly involved in processes, something that resonates well with design theory.

Below we develop the pragmatist concepts of *situation, inquiry* and *technology* into an analytical frame that enable us to address how design materials can support ideation.

## Problematic situations

From a Deweyan perspective, we are always engaged in a rhythmic exchange with our environment (Dewey 1934). Existence manifests as experience, most of which most do not stand out and give rise to inquiry. However, at times we find ourselves in *problematic situations* – something in our environment is not satisfactory, hard to understand, unresolved or intriguing. Thus we are always in transaction with our environment – when a problematic situation is recognized it is the fact that the rhythm of our existence is somehow out of tune with our surrounding environment. As we are just one of many such existing processes in a given situation, this out of tune-ness might be due to many different changes, but the main point being that there is now a need to recover the balance with our immediate environment. In life in general and in design in particular we might place ourselves in such situations of resistance by choice (Gedenryd 1998b). Designers know that design is a *process* that requires them to make choices in situations which does not have any right or wrong answers – design deals with wicked problems (Buchanan 1992). The idea of wicked problems resonates well with pragmatism, in that pragmatism does away with right or wrong answers as set categories. Rather a solution to a problematic situation can be good or bad, dependent on whether it helps restore the rhythm with the immediate environment.

## Inquiring strategies

*Inquiring strategies* are the approaches taken towards resolving a problematic situation. When problematic situations occur we enter into a mode of *inquiry* in which we, through exchange with our environment, try to resolve the problematic situation through a gradual and experimental transformation of the situation's constituent elements (Dewey 1925). By doing so we progress "through" a situation by applying *inquiring strategies*. In a pragmatist view this approach consists both of what you do as well as the rationale for it. We change our environment in order to satisfy and resolve whatever problematic situation we have encountered by re-establishing the functional coordination *(Garrison 2009)* with our surroundings. In design this resolving of unsatisfying situations encompass both design thinking and action. Design thinking (Cross 2006) or design judgement (Nelson and Stolterman 2014) means, in a pragmatist perspective, reformulating one's appreciation of a problematic situation, seeing and appreciation the situation from a new perspective. Inquiry in a design thinking perspective is the appreciative part of transforming and understanding a problematic situation, as well as formulating hypothesises for its resolution. However in a pragmatist understanding, inquiry might as well be aimed at external conditions – at functional coordination with ones physical surroundings. This is crucial to understanding the interplay of design materials and design thinking: In a pragmatist view of resolving problematic situations through inquiring strategies, any distinction between internal (intellectual) or external (physical) transformations is a purely methodological one, rather than a metaphysical one.

## Transformation

Inquiry unfolds through formations and tests of hypotheses – in this way problematic situations are transformed. In a Deweyan perspective, inquiry is supported by technology, which is the part of a situation used to transform a problematic situation. Although technology is a broad concept in pragmatism – it encompasses anything used for transforming a situation including theoretical constructs and tools – we focus on a particular kind of technology, materials used in creative design processes. In pragmatism, technology, and thereby materials constitute the problematic situation but are also at the same time part of the inquiry aimed at resolving the problematic situation. In action, materials give shape and direction to the problematic situation, but are also shaped and used in mediation inquiry. Gedenryd (Gedenryd 1998a) highlights how this can be seen as *making the world part of the cognition,* drawing our attention to how thinking and doing are intertwined. In our perspective, viewing materials as pragmatic technology allows us to analyze how materials help designers resolve problematic situations through inquiring strategies – we focus on which role specific materials (technology) played in different design phases (problematic situations).

## Methodology

The research conducted builds on a *research-through-design (Zimmerman, Forlizzi and Evenson 2007)* and has a particular focus on the design process (Basballe and Halskov 2012).

Our research laboratory is pursuing several research agendas (Halskov 2011). First we have a longstanding tradition for working with media architecture investigating technical as well as design process issues arising within this particular domain. Second we are generally interested in understanding creative leaps in the design process, in understanding how collaboration unfold, and how different design materials support collaboration and creativity. To investigate these questions we typically collaborate with industry partners in design processes in which we participate as designer-researchers interweaving research interests and design interests (Basballe and Halskov 2012).

One of our recent design cases is the design of the media architecture part of Danish pavilion at Expo 2010. In order to enable our research agendas, among these the role of design materials in throughout the design process, we carefully documented the design process an internally developed software tool (Dalsgaard and Halskov 2012). We documented workshops, on-going design work, technical issues and communication with our partners, the architect and the media architecture technology provider. The documentation of the process consists of both rich amounts of text, video and images, as well as the code repositories established throughout the process. By having a rich data material and reflecting on it through papers such as these, we are able to contribute to the massive body of research already outlined by (Wiberg 2013) by considering in detail the design process as a practice of crafting and re-crafting materials. To strengthen the arguments in the following analysis we examine our own design process in detail, specifically focused on three ensembles of materials: a *mockup*, a *pixel tool*, and a *mixed reality model*.

## The expo 2010 pavilion

The commission for the Danish pavilion at Expo 2010 was awarded to the Danish architectural firm, BIG, in a joint venture with the construction engineers at Arup. The outer facade of the pavilion is perforated with approximately 3.600 holes of various sizes and configurations.

Our research laboratory became involved in the project by the time the design of the building was already determined; the original idea was that the holes would simply be plain holes. We had previously collaborated with BIG in the field of media facades, and with the Danish lighting manufacturer, Martin Professional, therefore the idea of turning the perforated facade into a media facade emerged quite naturally.

The idea of illuminating the nearly four thousand holes in the facade presented a good case for a media facade that would articulate the expressive pattern of the facade during the evening hours of the Expo. The basic idea was to add lighting fixtures to the cavity wall, above each tube passing through the facade. The tube would then have to be made from a semi-translucent material that would make each hole appear as an illuminated, tube-shaped pixel on a media wall. This meant that the lighting fixtures would be hidden, thereby becoming a part of the building.

## Designing the Expo 2010 Pavillion

Here we analyse the design process with a focus on the materials used in three specific episodes during the design process: *mockup, pixel tool* and *mixed reality model*. Materials play a key role in each of these three episodes as tools for thinking and advancing the design process.

### The mockup

The mockup was a full-scale, wood model of a section of the facade (see fig x) built by BIG. We used the mockup to test light fixtures and the quality, colors and intensity of the individual "pixels".

The *problematic situation* at this point in the design process concerned the holes as pixels: how might we make the lights appear as pixels; how would we get the most useful appearance of pixels for a media facade; and how would we mount the light fixtures and lights? To resolve these questions we needed to experiment and explore the possibilities and constraints offered by tubes, lights and fixtures. For instance, since the lights were placed inside a tube, there were doubts whether the distribution of the lights would have a uniform appearance. While these issues might seem to be somewhat technical issues, we knew that the pixel appearance would greatly influence our options for later designs - in that sense the problematic situation here was an explicit exploration of one part of the project design space in order to set the stage for later developments.

We approached the situation with an *inquiring strategy* of experimenting with the mockup to inquire into how and whether creating individual pixels using lights and fixtures was possible. We start our inquiry by asking broad questions, and work our way forward through inquiry using the design materials. We experimented with how different kinds of light fixtures could be mounted behind various kinds of PVC tubes while pursuing the goal of achieving a uniform distribution of light. These experiments alternated between the details of the design materials (how will this fit here, how much light comes out) and the whole (can we have pixels that turns on and off etc.). The light fixtures could easily be combined with the wooden structure of the mock-up and we were free to experiment with different ways of combining the materials of light fixtures tube and the wooden structure. By using the mockup to support our inquiring strategy we were able to move inquiry forward, asking whether the idea of "holes in the buildings as pixels" would work.

Our work with the mockup illustrates how *design materials supported transformation.* Since the mockup consisted of materials that we could easily transform we were able to ask general questions as well as detailed technical questions, thus transforming problematic situation and moving inquiry forward. Each question was formulated by us, and then asked through experimenting with the design materials. Each experiment with the mockup is a transformation of the design materials, just as each new insight gained by doing so was a transformation of our understanding. Having a mockup made of wood means that the transformation of the materials was easy to accomplish - and each transformation contributed to moving inquiry forward towards the resolution of the problematic situation. While the mockup was made of wood, a material that could easily be shaped, the lights and fittings were the materials BIG wanted to use in the finished Expo2010 facade and were not as easily transformed. The design materials of the mockup thus supported transformation in two ways - first by helping us gain a very concrete understanding of the issues and potentials at hand thus establishing the initial problematic situation. And second by offering up potential transformations that helped move inquiry forward, both through changing the design and by helping us transforming our own understanding.

In this way we gradually transformed this initially unstable and doubtful situation into a working prototype with light fixtures behind PVC tubes turned into pixels.

## Pixel tool

The pixel tool was a Flash-based software application able to visualize a small section of twenty-four of the total 627 columns in the façade (Figure xx). The facade section was 2-dimensional and the "pixels" shown in the pixel tool approximated the tubes as circles and crescent shapes, depending on the perspective. The pixel tool would map video content onto the pixel shapes, thus giving us an idea of how different video content would appear on the finished facade.

In this phase of the design process the *problematic situation* started from our experience that patterns and shapes of pixels would impose challenges and limitations on the possible the video materials displayed on the facade could take. While the mockup dealt with one aspect of the design process, the actual shape and appearance of the pixels, we were here faced with a question of what capabilities for showing digital materials the finished facade would have.

Our *inquiring strategy* consisted of building the pixel tool and testing a wide range of visual content, including stock video footage, graphical animations, text, gradients, and abstract graphical patterns. Each piece of digital material tested in the pixel tool contributed to advancing inquiry. By allowing us to experiment with different materials the pixel tool allowed us to both try out finished digital materials (stock photos, animations from previous projects) as well as materials developed for the specific purpose of displaying on this media facade. It was quickly revealed that content that worked for regular screens and higher resolution, uniformly scaled media facades, was not possible to use here. Inquiry unfolded by experimenting with combining two materials - the pixels on the facade as well as the many potential digital materials. Through experimentation with the pixel tool and different content we realized that not only the low resolution posed problems, but in particular the lack of clear horizontal lines in the facade made traditional geometric figures hard to perceive and text only faintly perceivable. The solution we arrived was to slow the speed of the content and use only very simple graphical content, designed to be very clear - otherwise viewers of the facade would not be able to make sense of the facade as screen.

In the case of the pixel tool this design material supported *transformation* by helping us develop a deeper appreciation of the materials we were working with. While some of the video material had been used in previous projects, the example of the pixel tool reveals how our understanding of the digital materials was situational. The materials that had worked in previous projects were beautiful and useful, but combined with the abilities and constraints of the Expo2010 facade these video materials would not work. This appreciation of materials in our work with the pixel tool highlights the interplay of transformations of materials and problematic situations: each experiment with a piece of digital content in the pixel tool advanced our understanding and informed our next design move. The pixel tool thereby formed a bridge between the digital materials and a software approximation of how the final facade would look. Although 2D, the pixel tool still supported transformations supporting our inquiring strategy as we gained an appreciation of how materials could be combined in the final project.

In sum, the pixel tool proved useful for exploring the not only the low resolution and pixel pattern but also an approximation of the pixel shape.

## Mixed reality model

The mixed reality model was a 1:100 scale physical model, onto which we projected the exact pixel configuration using two video projectors (Dalsgaard and Halskov 2011). Using virtual 3D technology, we were able to show the holes as they would be illuminated on the pavilion and simulate the light and shadows cast by the sun. By inputting video material into the virtual 3D model we were thus able to experiment with the combination of physical scale model and digital video material.

The *problematic situation* in this example concerned how different kinds of content would appear on the entire curved 3D shape (as opposed to the sectional 2Dform explored by the pixel tool). Many of the previous experiments in the design process explored aspects of this part of the design space. However the mixed reality model represented our first attempt at bringing the digital material (in the form of video and 3D) together with the physical (in the form of the curved 3D shape of the facade). The problematic situation concerned bringing together the physical with the digital, and evolved to being concerned with how we might develop different specific digital video materials for use on the final Expo2010 platform.

The mixed reality model was part of our *inquiring strategy* to resolve the problematic situation of how digital materials would appear on the curved 3D form. We had previously created a 3D model that was imported into Unity, a 3D engine. By using projectors to combine this 3D model with a physical scale model, we could simulate both the curved form of the projected building and the digital materials projected onto it. Inquiry developed by trying out different materials in the 3D model and projecting them onto the physical model. This work gave us an idea of how different materials would appear and behave on the finished media facade. The mixed reality model allowed us to resolve the problematic situation discussed above, by facilitating a series of experiments with different digital materials on the already decided physical facade.

By allowing us to experiment with the combination of different materials, the mixed reality model supported *transformations* of our thinking as well as the proposed design solution. By allowing us to see our video material in the context of the curved 3D shape, we were able to revisit the strengths and weaknesses previously considered with the pixel tool. That led to transformations of our understanding and appreciation of the capabilities of the facade, driving inquiry forward. Because the mixed reality model was built with development in mind, it also supported external transformations of the problematic situation. We were able to load new video material into the 3D model, and project it onto the scale model, meaning that experimentation with different materials was fast and easy. By jumping back and forth between trying out new video materials and examining their appearance on the mixed reality model, design materials supported our inquiry through both internal and external transformations.

# Discussion

We have conceptualized design materials as pragmatic technology to build a framework suitable for analyzing the role of design materials in design processes. Our works extends Schön conceptualization of design as a conversation with a situation (Schön 1992) by developing Dewey's pragmatic philosophy. Specifically we build on the pragmatist concepts of situation, inquiry, transformation (Dewey 1925, 1934). Our emphasis here has been on material from a pragmatist perspective: how does design materials 1) establish *problematic situations*, 2) help resolve them through *inquiring strategies,* and 3) thus support *transformation* of the problematic situation. Below we discuss these three roles for materials.

## Establishing problematic situations

In creative design processes designers explore design questions. Materials are part of this exploration by externalizing and representing parts of a design. Looking back at our three examples above, they all examine different aspects of designing the Expo 2010 facade. In the case of the mock-up, we used physical materials (wood, light fixtures and lights) to externalize a specific part of the overall design, asking how might we use the holes in the facade as pixels? In the case of the pixel tool, we externalized another part of the overall design, by asking what capabilities for showing digital content part of the finished facade would have. With the mixed reality model, we posed the question of how different digital video materials would combine with the curved 3D shape of the facade.

In all three examples, we used materials to establish a problematic situation. We started out from an overall problematic situation of asking how we might design the Expo 2010 facade as a media facade. Throughout the design process, (as illustrated in each of the analyzed examples) we jumped between the details of the problematic and the overall design aims. Materials played a key role here - in each of the examination of part of the overall problematic situation, we used materials to constitute part of the design space we were interested in during the particular phase. Thus, materials helped us externalise part of a design space, allowing us to develop inquiring strategies to resolve the problematic situation.

### Developing inquiring strategies

Materials did not just establish problematic situations in a tangible form - they also helped develop inquiring strategies for resolving the problematic situations. In design, problems and solutions co-evolve, and in our pragmatist framework, this is reflected in the relationship between *problematic situations* and *inquiring strategies*. Design materials serve a dual role - they both at the same time establish a problematic situation and offer the potential for transforming the problematic situation. Thus, using different materials to establish a design situation, also offer different potential ways of viewing and resolving it.

In our analysis this was highlighted by using the materials of the mockup to move forward by exploring the quite technical question of creating pixels using the holes, light fixtures and lights. In the case of the pixel tool, we used digital materials to experiment with a 2D prototype of the physical facade. Experimental use of the pixel tool revealed new qualities of the facade showing how inquiring strategies was supported by materials. By establishing the problematic situation and developing inquiring strategies using materials.

### Supporting transformations

Inquiry is the directed and experimental transformation of a problematic situation. Transformations are always supported by technology, and the mockup, pixel tool and mixed reality model all supported us in transforming the problematic situations. For instance, the mockup helped us establish and transform the problematic situation. The mockup did this by having the actual light fixtures and a wooden shape of the holes, meaning that we were able to first transform the mockup, then transform our own perception of the what was possible to achieve using the holes and lights to create pixels. That highlights how transformation of design questions can be directed at internal as well as external parts of the problematic situation. Transformations, internal or external, means gradually and experimentally establishing and developing problematic situations and inquiring strategies. In the same way, the mixed reality model supported transformations by letting us alternately test new video materials and examine their appearance on the mixed reality model, thus letting our transformations of the mixed reality model move inquiry forward by offering transformations of our perception of problematic situation.

# Conclusion

In this paper we have been in developing a theoretical frame for understanding design materials in a pragmatist perspective. Rather than analyze the entire Expo2010 design process, we have chosen three examples that are complex enough to allow us to reflect in depth on design materials and their role in the design processes. While the mockup examines a single pixel in physical materials, the pixel tool is a digital investigation of a 2D facade, and we end up discussing the mixed reality model, which combines 3d models on a curved 3d shape. Thus in the three examples we examined different aspects of the Expo2010 facade using different materials. Each example contributed to moving the process forward by allowing us to jump between the overall design idea and different details.

By making the case that materials can be understood as *pragmatic technology (Hickman 1992)* we offer a way of understanding the role of materials in *establishing problematic situations, developing inquiring strategies* and *supporting transformations*. In all three roles, materials are seen as something that offers specific constraints and possibilities for perceiving and changing the situation until it is, in pragmatist terms, satisfactory. Understanding design materials as pragmatic technology draws attention to how we use design materials to explore and resolve specific problematic situations. And in an even loftier perspective we might understand design processes as the use of design materials to understand, explore and resolve wicked problems.